\documentclass[traditabstract]{aa} 

\usepackage{graphicx}
\usepackage{txfonts}
\usepackage{longtable}
\usepackage{lscape}
\usepackage{hyperref}
\hypersetup{colorlinks=true, citecolor=blue}
\usepackage{amstext}

\begin{document}

   \title{A Virgo Environmental Survey Tracing Ionised Gas Emission (VESTIGE)\\ XV. The H$\alpha$ luminosity function of the Virgo cluster\thanks{Based on observations obtained with
   MegaPrime/MegaCam, a joint project of CFHT and CEA/DAPNIA, at the Canada-French-Hawaii Telescope
   (CFHT) which is operated by the National Research Council (NRC) of Canada, the Institut National
   des Sciences de l'Univers of the Centre National de la Recherche Scientifique (CNRS) of France and
   the University of Hawaii.}
      }
   \subtitle{}
  \author{A. Boselli\inst{1,**},  
          M. Fossati\inst{2,3},
          P. C{\^o}t{\'e}\inst{4},
          J.C. Cuillandre\inst{5},
          L. Ferrarese\inst{4},
          S. Gwyn\inst{4}, 
          P. Amram\inst{1},
          M. Ayromlou\inst{6},
          M. Balogh\inst{7},
          G. Bellusci\inst{2},
          M. Boquien\inst{8},
          G. Gavazzi\inst{2},
          G. Hensler\inst{9},     
          A. Longobardi\inst{2},
          D. Nelson\inst{6},
          A. Pillepich\inst{10},
          J. Roediger\inst{4},
          R. Sanchez-Jansen\inst{11},
          M. Sun\inst{12},
          G. Trinchieri\inst{3}
       }

\institute{     
                Aix Marseille Univ, CNRS, CNES, LAM, Marseille, France\thanks{Scientific associate INAF - Osservatorio Astronomico di Brera, via Brera 28, 20159 Milano, Italy}
                \email{alessandro.boselli@lam.fr}
        \and
                Universit\'a di Milano-Bicocca, piazza della scienza 3, 20100 Milano, Italy
        \and
                INAF - Osservatorio Astronomico di Brera, via Brera 28, 20159 Milano, Italy
        \and
                National Research Council of Canada, Herzberg Astronomy and Astrophysics, 5071 West Saanich Road, Victoria, BC, V9E 2E7, Canada
        \and
                AIM, CEA, CNRS, Universit\'e Paris-Saclay, Universit\'e Paris Diderot, Sorbonne Paris Cit\'e, Observatoire de Paris, PSL University, F-91191 Gif-sur-Yvette Cedex, France
        \and    
                Universit\"at Heidelberg, Zentrum f\"ur Astronomie, Institut f\"ur theoretische Astropysik, Albert-Ueberle-Str.2, 69120 Heidelberg, Germany
        \and
                Waterloo Centre for Astrophysics, University of Waterloo, Waterloo, Ontario N2L3G1, Canada        
        \and
                Centro de Astronomi\'a (CITEVA), Universidad de Antofagasta, Avenida Angamos 601, Antofagasta, Chile
        \and
                Department of Astrophysics, University of Vienna, T\"urkenschanzstrasse 17, 1180 Vienna, Austria
        \and
                Max-Planck-Institut f\"ur Astronomie, K\"onigstuhl 17, 69117 Heidelberg, Germany
        %
        \and
                STFC UK Astronomy Technology Centre, The Royal Observatory Edinburgh, Blackford Hill, Edinburgh, EH9 3HJ, UK
        \and
                Department of Physics \& Astronomy, University of Alabama in Huntsville, 3001 Sparkman Drive, Huntsville, AL 35899, USA
                 }

\authorrunning{Boselli et al.}
\titlerunning{VESTIGE}

   \date{}

 
  \abstract  
{We use a complete set of deep narrow-band imaging data for 384 galaxies gathered during the Virgo Environmental Survey Tracing Ionised 
Gas Emission (VESTIGE) to derive the first H$\alpha$ luminosity function of the Virgo cluster within its virial radius. The data, which 
are sensitive to the emission of a single O-early B ionising star, allow us to cover the whole dynamic range of the H$\alpha$ luminosity 
function (10$^{36}$ $\leq$ $L(H\alpha)$ $\leq$ 10$^{42}$ erg s$^{-1}$). After they are corrected for [NII] contamination and dust attenuation, 
the data are used to derive the star formation rate function in the range 10$^{-4}$ $\lesssim$ $SFR$ $\lesssim$ 10 
M$_{\odot}$ yr$^{-1}$. These luminosity functions are derived for gas-rich 
and gas-poor systems and for objects belonging to the different substructures of the Virgo cluster. They are then compared to those 
derived at other frequencies or using different tracers of star formation in Virgo, in other nearby and high-$z$ clusters, in the 
field, and finally to those predicted by the IllustrisTNG cosmological hydrodynamical simulations (TNG50 and TNG100). 
The H$\alpha$ luminosity function of the Virgo cluster is fairly flat ($\alpha$ = -1.07 when fitted with a Schechter function) 
in the range 10$^{38.5}$ $\lesssim$ $L(H\alpha)$ $\lesssim$ 10$^{40.5}$ erg s$^{-1}$, and it abruptly decreases at lower luminosities.
When compared to those derived for other nearby clusters and for the field, the slope and the characteristic luminosity of the 
Schechter function change as a function of the dynamical mass of the system, of the temperature of the X-rays gas, and of the
dynamical pressure exerted on the interstellar medium of galaxies moving at high velocity within the intracluster medium. All these trends can be explained 
in a scenario in which the activity of star formation of galaxies is reduced in massive clusters due to their hydrodynamical interaction
with the surrounding medium, suggesting once again that ram-pressure stripping is the dominant mechanism affecting galaxy
evolution in local clusters of dynamical mass $M_{cluster}$ $\gtrsim$ 10$^{14}$ M$_{\odot}$. The comparison with the 
IllustrisTNG cosmological hydrodynamical simulations shows a more pronounced decrease at the faint end of the distribution. If the Virgo 
cluster is representative of typical nearby clusters of similar mass, this difference suggests that the stripping process in simulated 
galaxies in these environments is more efficient than observed.
}
   {}
   {}
   {}
   {}
   {}


   \maketitle
%

\section{Introduction}

The properties of galaxies inhabiting rich environments are significantly different than those in the field. They are principally relaxed systems
(elliptical and lenticulars; Dressler 1980, 1997), while the late-type population is characterised by a reduced atomic (Haynes \& Giovanelli 1984; 
Solanes et al. 2001; Gavazzi et al. 2005) and molecular (Fumagalli et al. 2009; Zabel et al. 2022) gas content. Because of this
lack of gas, cluster galaxies also have a reduced star formation activity compared to that of similar objects in the field 
(e.g. Gavazzi et al. 1998; Lewis et al. 2002; Gomez et al. 2003; Peng et al. 2010; Boselli et al. 2014).

Different perturbing mechanisms have been proposed to explain these systematic differences between cluster and field galaxies.
They include gravitational perturbations with nearby companions (e.g. Merritt 1983), with the gravitational potential well of the 
cluster itself (Byrd \& Valtonen 1990),
and their combined effect in multiple fly-by encounters (galaxy harassment; Moore et al. 1998), the hydrodynamic interaction of
the interstellar medium (ISM) of galaxies with the surrounding hot ($T$ $\simeq$ 10$^7$-10$^8$ K) and low-density ($\rho_{IGM}$ $\simeq$ 10$^{-3}$ cm$^{-3}$)
intergalactic medium (IGM) (ram-pressure stripping, Gunn \& Gott 1972, Boselli et al. 2022a; thermal evaporation; Cowie \& Songaila 1977;
viscous stripping; Nulsen 1982), and the gentle stop of gas infall on the disc when galaxies become satellites of larger haloes 
(starvation; Larson et al. 1980). These perturbing mechanisms have different effects on the process of star formation. If the lack of gas
that is  removed during all types of interactions leads to a reduction of the star formation activity, this quenching phenomenon occurs on 
different timescales. In a starvation scenario, the cessation of the infall of fresh gas induces a slow reduction of the star formation activity 
on the disc that fades as soon as the available cold ISM is consumed by star formation (e.g. Larson et al. 1980, Boselli et al. 2006).
In gravitational perturbations, which are able to simultaneously remove all the baryonic components of galaxies (gas, dust, and stars), the instabilities
created during the encounter induce gas infall into the nucleus, possibly feeding a centralised starburst activity while decreasing star formation in the disc
(e.g. Lake et al. 1998). Statistically, the typical timescales necessary
to reduce the activity of star formation of harassed galaxies are also very long because multiple fly-by encounters with other cluster
members are necessary (eg. Moore et al. 1998; Gnedin et al. 2003; Bialas et al. 2015). Hydrodynamic 
interactions, which act only on the different phases of the ISM, remove the gas from the outside in, 
producing truncated gaseous and star-forming discs (see for a review Boselli \& Gavazzi 2006, 2014). There is growing evidence from observations
(Boselli et al. 2016a; Fossati et al. 2018) and simulations (Quilis et al. 2000; Schulz \& Struck 2001; Tonnesen et al. 2007; 
Roediger \& Bruggen 2007; Lotz et al. 2019) that these mechanisms can remove the gas and reduce the activity of star formation
of the perturbed galaxies on timescales shorter ($\lesssim$ 0.5-1 Gyr) than the typical cluster-crossing timescale ($\simeq$ 1.7 Gyr).
The quenching process is particularly efficient in dwarf systems, where the gas is loosely bound to the shallow gravitational 
potential well of the galaxy (e.g. Boselli et al. 2008).

The analysis of different sets of data aimed at identifying the dominant perturbing process, 
combined with tuned models and simulations, still gives inconsistent results. While the analysis of nearby 
samples of galaxies with multifrequency data (e.g. Vollmer et al. 2001; Gavazzi et al. 2010; Boselli et al. 2014;
Yagi et al. 2010) and the detailed analysis of selected representative objects (e.g. Vollmer et al. 2008; 
Kenney et al. 2014; Fossati et al. 2018; Longobardi et al. 2020; Boselli et al. 2021; Junais et al. 2021; 
see also Poggianti et al. (2017) and the GASP survey\footnote{GAs-Stripping Phenomena in galaxies with MUSE}) 
all indicate ram-pressure stripping 
as the dominant perturbing mechanism in nearby clusters such as Coma, Virgo, and A1367 (Sun et al. 2006; 
Zhan et al. 2013; Merluzzi et al. 2013; Jachym et al. 2014; Fumagalli et al. 2014; Fossati et al. 2016; Gullieuszik et al. 2017; 
Bellhouse et al. 2019; Ramatsoku et al. 2019). Other statistical studies based on large samples
(Wetzel et al. 2013) or on selected regions of the sky (e.g. McGee et al. 2009; 
De Lucia et al. 2012; Haines et al. 2015) rather favour long timescales for quenching
and so possibly starvation. These inconsistent results, as extensively discussed in Boselli et al. (2016a), can be explained by several facts, including 
the use of different age-sensitive tracers, of different representative samples, or possibly of inconsistent definitions 
of cluster and/or satellite galaxies in the various observations and simulation analyses.

The use of multifrequency observations combined with tuned models and simulations is of paramount importance for identifying the dominant perturbing
mechanism in rich environments. Of particular importance are all age-sensitive tracers that can be used to quantify the typical timescales 
for reducing the star formation activity of the perturbed galaxies. A particular role is played by the H$\alpha$ line emission, which
is the only tracer able to resolve quenching episodes on timescales shorter than 10 Myr (Kennicutt 1998). 

With the purpose of studying the effects of the perturbations on the star formation activity of cluster galaxies, we are undertaking
the Virgo Environmental Survey Tracing Ionised Gas Emission (VESTIGE; Boselli et al. 2018a). This is a deep H$\alpha$ narrow-band imaging survey
of the whole Virgo cluster up to its virial radius, corresponding to 104 deg.$^{2}$ on the plane of the sky. The excellent quality of the data in terms of sensitivity and angular resolution
allowed us to make detailed studies of several representative objects undergoing a perturbation (e.g. Boselli et al. 2018b;
Fossati et al. 2018; Vollmer et al. 2021; Sardaneta et al. 2022) and to analyse the quenching process 
down to the scale of individual HII regions (Boselli et al. 2021). VESTIGE is now almost complete (76\%) and allows us to 
make the first statistical studies on the H$\alpha$-derived star formation properties of the perturbed galaxies. We have recently completed
a dedicated work on the origin of the ultra-diffuse galaxies inhabiting the cluster (Junais et al. 2022) and another study on the
main-sequence relation (Boselli et al. 2023). In this work, we present the H$\alpha$ luminosity function of Virgo. 
The luminosity function of galaxies is a statistical tool that is often used in the literature to constrain numerical models of galaxy evolution
in the full cosmological context.
It has been used, for instance, to show the different slopes observed at the faint and bright ends between observed 
galaxy populations and the predictions of simulations and semianalytic models to prove the importance of the feedback mechanism by supernovae and 
active galactic nuclei (AGNs)
in galaxy evolution (e.g. Benson et al. 2003; Croton et al. 2006; Somerville \& Dave 2015). In the cluster environment, the optical luminosity 
function has shown an excess of dwarf galaxies with respect to the field (Yagi et al. 2002; de Propris et al. 2003; Popesso et al. 2005; Ferrarese et al. 2016;
see however Blanton et al. 2005). The origin of this difference is still debated.

Constructing the H$\alpha$ luminosity function of nearby clusters of galaxies is particularly challenging for several reasons. 
The first reason is that it requires an H$\alpha$ selection that can only be made when untargeted surveys are available.
At present, this can be done only using two different techniques. The first technique is based on objective prism spectroscopy, and the 
second technique uses narrow-band imaging. The angular extension on the sky of nearby clusters means that wide-field cameras and prisms are required, 
while large apertures are required to achieve the sensitivity required to detect the weak emission of quenched systems. Slitless grism spectroscopy,
now often used for environmental studies at high redshift with the Hubble space telescope (HST; e.g. Zeimann et al. 2013; Vulcani et al. 2017;
Fossati et al. 2017; Matharu et al. 2022), has been used in the past for a few nearby clusters 
using photographic plates with the Burrell-Schmidt telescope at Kitt Peak (Moss et al. 1988; Moss \& Whittle 2000). These works
were limited to fluxes of $f(H\alpha)$ $\simeq$ 10$^{-13}$ erg s$^{-1}$ cm$^{-2}$ and equivalent widths of $\simeq$ 20 \AA,\, that is, to the 
most active galaxy population. Wide-field narrow-band imaging has become available at several 4- to 8-metre class telescopes 
only in recent years. Despite this, studies of nearby clusters are still limited principally because 
very expensive large narrow-band filters especially designed to match their transmissivity curve with the velocity distribution 
of the target clusters are lacking. The only exceptions are the clusters A1367 and Coma, which were studied in detail with the Subaru telescope (Yagi et al. 2010, 2017).
These works provided spectacular results, such as the identification of extended low surface brightness ionised gas tails
associated with $\simeq$ 50\%\ of the late-type galaxies, proving that ram-pressure stripping is the dominant perturbing mechanism here.
The data obtained so far, however, still did not allow us to derive the H$\alpha$ luminosity function of the two clusters because of several main 
technical problems, such as the subtraction of the stellar continuum and strong reflection of bright stars, which hamper
a correct measure of the total flux of the detected sources. Narrow-band imaging has been used to measure the bright end ($L(H\alpha$ $\gtrsim$ 10$^{40}$-10$^{41}$ erg s$^{-1}$)
of the H$\alpha$ luminosity function of a few clusters at higher redshift (0.4 $\lesssim$ $z$ $\lesssim$ 0.8; e.g. Kodama et al. 2004; Koyama et al. 2010; Sobral et al. 2016). 
Even though a similar observational technique was adopted, the results obtained so far are not always consistent: Some clusters are characterised by a steep rise at faint luminosities (Kodama et al.
2004; Koyama et al. 2010), and others by a relatively flat slope in the densest regions (Sobral et al. 2011).

VESTIGE provides us with a unique set of H$\alpha$ narrow-band imaging data that is perfectly suited for this purpose. It is the widest 
and deepest survey for which this exercise can be performed. Its sensitivity is five orders of magnitudes better than that of
previous prism surveys. The sensitivity is comparable to that gathered with Subaru for Coma and A1367 because of the excellent transmissivity of the narrow-band filter, the long exposures, and the observing strategy, which is especially tailored to optimise the flat fielding (see Sec. 2) so that the low surface brightness tails of stripped ionised gas can be detected and the flux of all detected galaxies can be measured simultaneously with exquisite accuracy ($<$ 10\%).
Furthermore, the proximity of the cluster (16.5 Mpc, Mei et al. 2007) allows us to detect
dwarf systems and thus to sample the luminosity function down to the emission of a single ionising O or early-B star.
The H$\alpha$ luminosity function presented in this work is thus the deepest ever achieved for any type of environment.
Although it is limited to a single cluster, it will be an excellent reference for years to come for cosmological galaxy simulations, which are now able to trace the
star formation activity down to the dwarf population.

The paper is structured as follows: In Sec. 2 we present the observations and the data reduction. In Sec. 3 we derive the 
star formation rate (SFR) of the detected galaxies after correcting the data for [NII] contamination and dust attenuation. 
In Sec. 4 we measure the H$\alpha$ luminosity function and SFR functions, fit them with a Schechter function, and
compare them with those derived for the field or for other clusters at different redshift. We discuss the results 
in the framework of galaxy evolution in Sec. 5 specifically in comparison to results from the Illustris TNG (The Next Generation) 
project, and we give our conclusions in Sec. 6.
 
\section{Observations and data reduction}

\subsection{VESTIGE narrow-band imaging}

The data analysed in this work were extracted from the VESTIGE survey. This H$\alpha$ narrow-band (NB) imaging survey
covers the whole Virgo cluster up to its virial radius (104 deg.$^2$). A full description of the data acquisition and reduction is given 
in Boselli et al. (2018a). Briefly, the data were collected using MegaCam at the CFHT
in two filters, the NB filter MP9603 centred on H$\alpha$ ($\lambda_c$ = 6591 \AA; $\Delta \lambda$ = 106 \AA), and the 
broad-band $r$ filter, which is necessary for the subtraction of the stellar continuum.
The NB filter is optimally designed to include the emission of the H$\alpha$ ($\lambda$ = 6563 \AA) and of the two [NII] lines
($\lambda$ = 6548, 6583 \AA) for all galaxies at the redshift of the cluster (-300 $\leq$ $v_{hel}$ $\leq$ 3000 km s$^{-1}$;
hereafter, we refer to the H$\alpha$+[NII] band as H$\alpha$ unless stated otherwise. 
The observations were carried out with two-hour integration in the NB filter and  with 12 minutes in the broad-band $r$ filter.
This last is necessary to secure the subtraction of the  stellar continuum (Boselli et al. 2019). 
The survey is 76\% complete as of February 2023. Full sensitivity has been reached almost everywhere within the VESTIGE footprint, 
with possible shorter exposures at the cluster periphery, where the survey still needs to be completed. The sensitivity of the survey in the NB
is $f(H\alpha)$ $\simeq$ 4 $\times$ 10$^{-17}$ erg~s$^{-1}$~cm$^{-2}$ (5$\sigma$) for point sources and $\Sigma(H\alpha)$ $\simeq$ 2 $\times$ 10$^{-18}$ erg~s$^{-1}$~cm$^{-2}$~arcsec$^{-2}$
(1$\sigma$ after smoothing the data to $\simeq$ 3\arcsec resolution) for extended sources. The sensitivity might drop by a factor of $\simeq$ 1.5 in some of the outer cluster regions, in which mapping has been secured by $\text{about}$ half of the programmed exposures ($\simeq$ 1 h). This limiting sensitivity is still far below
the sensitivity that was used to constrain the $H\alpha$ luminosity function, and this incompleteness
therefore does not affect the results of this work.
The observations were carried out in excellent seeing conditions in both bands (FWHM=0.76\arcsec\ $\pm$ 0.07\arcsec).

The data were reduced using Elixir-LSB (Ferrarese et al. 2012). This pipeline is especially designed to optimise the detection 
of low surface brightness and extended regions such as those expected in galaxies interacting with their surrounding environment. The photometric
calibration and the astrometric corrections of the images were made using standard MegaCam procedures such as those described in Gwyn (2008).
The resulting photometric accuracy in the two bands is $\leq$ 0.02-0.03 mag.


\subsection{Galaxy identification}

The H$\alpha$ emitting sources we used in our analysis were identified as extensively described in Boselli et al. (2023).
Since completeness in the selection is necessary for an accurate determination of the luminosity function, we recall 
the adopted criteria here.
Galaxies were identified as line emitters in the continuum-subtracted images as
1) counterparts of the galaxies included in the Virgo cluster catalogue (VCC; Binggeli et al. 1985) and identified as Virgo cluster members
based on their redshift ($v_{hel}$ $\leq$ 3000 km s$^{-1}$; 307 objects);
2) ounterparts of sources detected in HI at the redshift of the cluster in the Arecibo Legacy Fast ALFA survey (ALFALFA; Giovanelli et al. 2005). ALFALFA
is a blind HI survey carried out at the Arecibo telescope and covers the whole footprint of VESTIGE, with a typical 
sensitivity of $rms$ = 2.3 mJy at 10 km s$^{-1}$ spectral resolution and 3.2\arcmin\ angular resolution. ALFALFA is thus able to detect sources 
at the distance of the Virgo cluster with HI masses down to $M_{HI}$ $\simeq$ 10$^{7.5}$ M$_{\odot}$. Because of the tight correlation between atomic gas 
content and star formation, HI emitting sources are ideal priors for H$\alpha$ detections.  
For this purpose, we searched for
any H$\alpha$ emitting counterpart at the position of the HI emitting sources in the Virgo cluster ($v_{hel}$ $\leq$ 3000 km s$^{-1}$) 
that were listed within the catalogue of Haynes et al. (2018) and identified 37 objects that were not included in the VCC.
3) counterparts of the 3869 galaxies identified as Virgo cluster members after the study of different scaling relations using the 
Next Generation Virgo cluster Survey (NGVS) data,
as described in Ferrarese et al. (2012, 2020) and Lim et al. (2020) (31 objects). This catalogue is complete to a stellar mass of 
$M_{star}$ $\simeq$ 1.6$\times$10$^7$ M$_{\odot}$, and it is 50\%\ complete to $M_{star}$ $\simeq$ 6.2$\times$10$^5$ M$_{\odot}$ (Ferrarese et al. 2020).\\
4) four bright galaxies with extended H$\alpha$ emission located outside the VCC and NGVS footprints.
5) five objects not included in the previous catalogues with clear extended line emission identified after visual inspection 
of all the VESTIGE continuum-subtracted images. Their fairly extended emission suggests that they are local low-mass blue compact dwarf (BCD) galaxies rather than
background line emitters. For two of them, we gathered a spectrum at the Apache Point Observatory (APO) 3.5~m telescope using the 
dual imaging spectrograph (DIS), and we identified (as expected)
a line emission within the NB VESTIGE band. The S/N in the spectrum, however, is too low to detect other emission or absorption lines, and thus
prevents an unambiguous identification of these sources as Virgo cluster members.  \\
Furthermore,
6) we excluded all point-source line emitters to avoid any possible contamination of background sources. Although their identification 
using multiple colour-colour diagrams derived combining the two VESTIGE bands (H$\alpha$ and $r$) with the four NGVS bands ($u, g, i, z$)
is still pending, a preliminary analysis of this sample suggests that they are predominantly Ly$\alpha$ emitters at $z$ $\simeq$ 4.4, as well as 
several [OII] and [OIII] emitters at $z$ $\simeq$ 0.8 and $z$ $\simeq$ 0.4, respectively. 
7) to statistically quantify the number of line emitters with extended emission, which thus might be potentially members of the Virgo cluster, and that 
might have been missed by this selection, 
we ran the code SExtractor (Bertin \& Arnouts 1996) in dual mode on the NB H$\alpha$ and on the broad-band $r$ images and identified
all resolved sources with an excess of emission in the NB (3$\sigma$). We then visually inspected all the sources with a radius $r$ $>$ 1\arcsec\
in both the NB and broad-band $r$ images and rejected all objects with a compact morphology ($r$ $\lesssim$ 3\arcsec) consistent with 
that of background galaxies at $z$ $\simeq$ 0.4 (e.g. van der Wel et al. 2014). We repeated this exercise on ten randomly
selected fields ($\simeq$ 10 deg.$^2$, $\simeq$ 10\%\ of the full survey), and we finally identified only one extended object composed of several 
compact star-forming regions, thus with a morphology which might suggest being member of the Virgo cluster. If it is assumed to lie at the distance of the cluster (16.5 Mpc), 
its luminosity would be $L(H\alpha)$ = 1.8 $\times$ 10$^{38}$ erg s$^{-1}$. Because this number is very small, we cannot apply any luminosity-dependent
statistical correction to the data.

To conclude, the sample of galaxies we used in this work to construct the H$\alpha$ and SFR luminosity functions is composed of 384 galaxies,
379 of which have a secure redshift identification. Because the NB filter  
transmissivity profile and the redshift distribution within the cluster match perfectly, we can consider the five remaining galaxies without a redshift (point 5 above) as highly probable Virgo cluster
members. The statistical properties of the sample, including the identification of galaxies in the phase-space diagram, are extensively described 
in Boselli et al. (2023). The sample spans the following ranges in H$\alpha$ luminosity: 10$^{36.5}$ $\lesssim$ $L(H\alpha)$ $\lesssim$ 10$^{42}$ erg s$^{-1}$,
corresponding to 10$^{-5}$ $\lesssim$ $SFR$ $\lesssim$ 10 M$_{\odot}$ yr$^{-1}$ in SFR (see Sec. 3.3), stellar mass, 
10$^6$ $\lesssim$ $M_{star}$ $\lesssim$ 10$^{11.5}$ M$_{\odot}$, and atomic gas mass, 10$^{6.5}$ $\lesssim$ $M_{HI}$ $\lesssim$  10$^{9.7}$ M$_{\odot}$.

\subsection{Flux extraction}

The flux extraction procedure is also extensively described in Boselli et al. (2023). It consists in measuring both the galaxy emission and the 
sky background on the continuum-subtracted images, which are calibrated with a photometric uncertainty of 
$\simeq$ 2-3 \%\ (see Sec. 2.1).
The galaxy emission and the sky background are measured within the same elliptical aperture defined to encompass all the emission associated 
with each galaxy, but at the same time, minimise the contribution of the sky background within the aperture. 
The ellipse includes only the galaxy emission, and it excludes any possible emission for extended features located outside the stellar disc
(extraplanar HII regions or diffuse gas tails), whose contribution is always minor. 
The flux and its uncertainty are then estimated by measuring the sky background 1000 times within about five times the diameter of the target
to minimise any possible effects due to large-scale residual gradients in the continuum-subtracted frame (the main source of uncertainty
in the estimated fluxes).
The uncertainties on the fluxes were obtained as the quadratic sum of the uncertainties on the flux counts and the 
uncertainties on the background (rms of the bootstrap iterations). The uncertainties on the flux counts were derived assuming 
a Poissonian distribution for the source photo-electrons. The uncertainty on the flux is generally $\lesssim$ 10\%. 
We also checked the quality of the flux calibration by comparing the
the H$\alpha$ fluxes derived in this work with those extracted from the integral field unit (IFU) spectroscopy gathered with 
the Multi-Unit Spectroscopic Explorer (MUSE) at the Very Large Telescope (VLT) for a few selected galaxies.
The accuracy is excellent in star-forming systems (three objects with a difference of $\leq$ 5\%; Boselli et al. 2018b, 2021, 2022b) and 
slightly lower in a bright lenticular (NGC 4526, $\simeq$ 18\%;\, Boselli et al. 2022b) because the subtraction of 
the stellar continuum emission in the NB data is difficulat.  

\subsection{Multifrequency data}

A large set of multifrequency data was used in the following analysis. These data are necessary for the identification of 
the H$\alpha$ emitting sources in the VESTIGE images (see Sec. 2.2) and for the correction of the H$\alpha$ data for [NII]
contamination and dust attenuation. Spectroscopic data were used to quantify the [NII] contamination and the Balmer decrement.
As in other VESTIGE papers, HI data were used to identify perturbed systems via the HI-deficiency parameter (Haynes \& Giovanelli 1984)
as those with $HI-def$ $>$ 0.4, where the HI-deficiency parameter is defined with the recent calibration of Cattorini et al. (2023). 
The complete set of multifrequency data is extensively described in Boselli et al. (2023).

\section{Derived parameters}

\subsection{[NII] contamination}

The [NII] contamination in the NB filter was derived using the spectroscopic data. In order of priority,
we used the VLT MUSE and the Focal Reducer and Low-dispersion Spectrograph (FORS) data whenever available (five galaxies; Fossati et al. 2018, Boselli et al. 2018b, 2021, 2022a),
the integrated spectra of the \textit{Herschel} Reference Survey (HRS; Boselli et al. 2015) (72 galaxies) and Virgo galaxies (Gavazzi et al. 2004) 
(44 objects), public Sloan Digital Sky Survey (SDSS) spectra (115 objects), or assuming the 
standard scaling relation between the [NII]/H$\alpha$ ratio and the stellar mass published in Boselli et al. (2023) (141 objects).

\begin{table}
\caption{Best-fit parameters for different H$\alpha$ luminosity functions}
\label{fit}
{
\[
\begin{tabular}{ccccc}
\hline
\noalign{\smallskip}
\hline
Variable                & sample                                & $\alpha$                      & log$L^*(H\alpha)$               & log$\phi^*$   \\
\hline  
$L(H\alpha)_{obs}$      & log$L(H\alpha)_{obs}$ $\geq$38        & -1.02$^{-0.98}_{-1.06}$       & 41.06$^{41.15}_{40.97}$ & 1.71$^{1.77}_{1.64}$\\
$L(H\alpha)_{[NII]}$    & log$L(H\alpha)_{[NII]}$ $\geq$38      & -1.02$^{-0.98}_{-1.06}$       & 40.85$^{40.94}_{40.76}$ & 1.75$^{1.82}_{1.67}$\\
$L(H\alpha)_{cor}$      & log$L(H\alpha)_{cor}$ $\geq$38        & -1.07$^{-1.04}_{-1.10}$       & 41.49$^{41.60}_{41.40}$ & 1.54$^{1.61}_{1.47}$\\
\noalign{\smallskip}
\hline
\end{tabular}
\]
Notes: units are erg s$^{-1}$ for the H$\alpha$ luminosities.
}
\end{table}

\begin{figure*}
\centering
\includegraphics[width=0.49\textwidth]{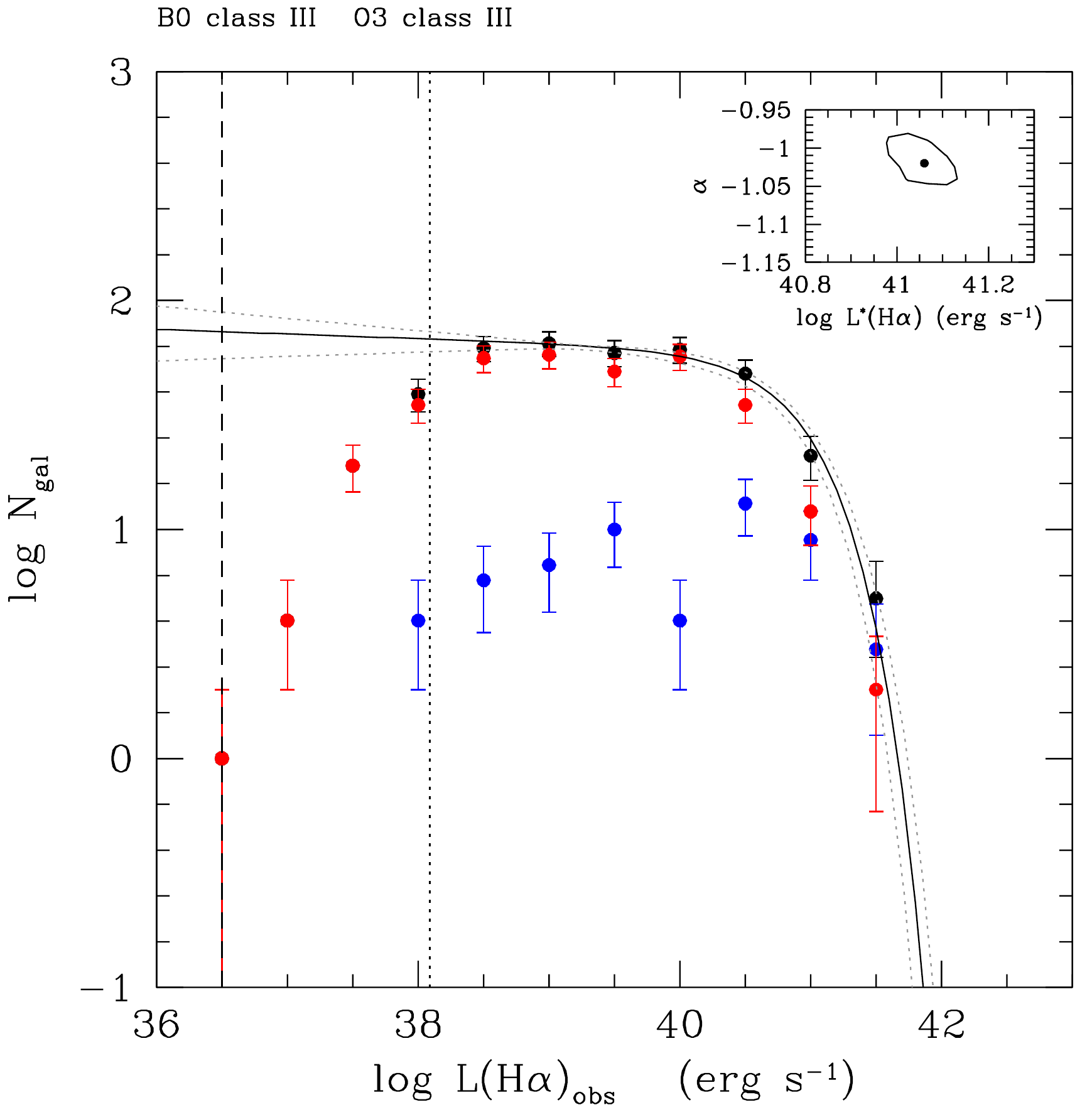}
\includegraphics[width=0.49\textwidth]{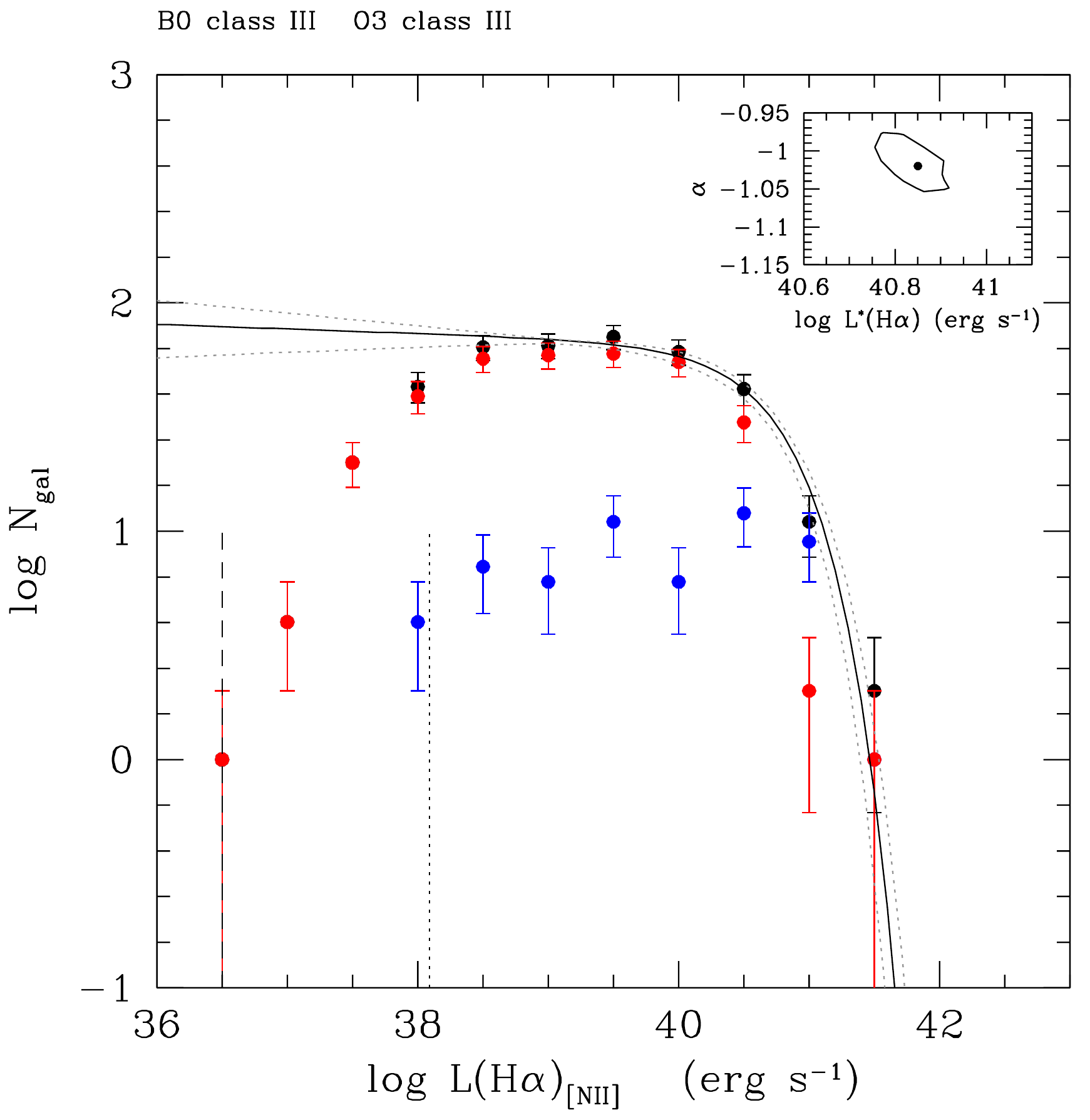}\\
\includegraphics[width=0.49\textwidth]{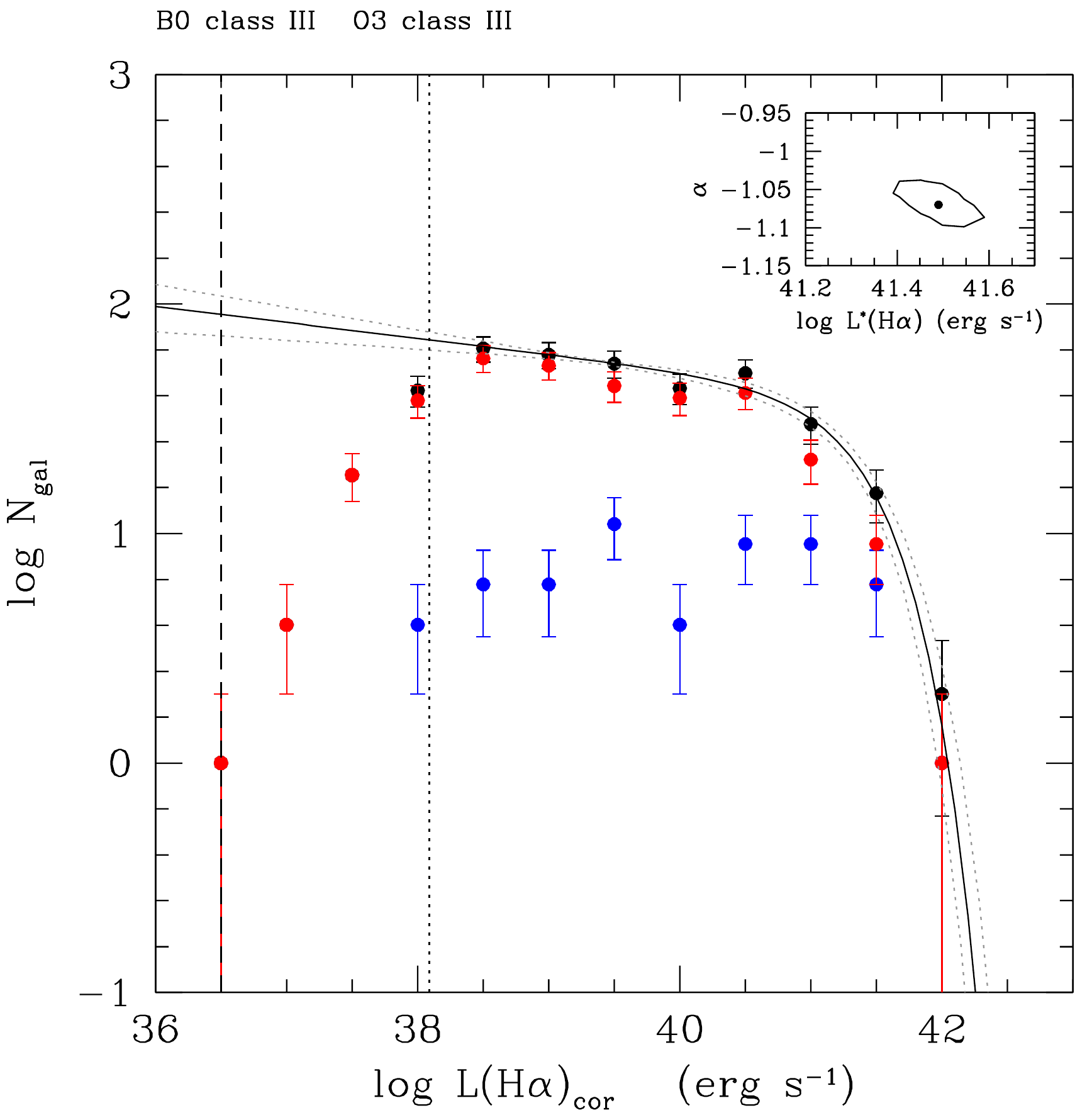}
\includegraphics[width=0.49\textwidth]{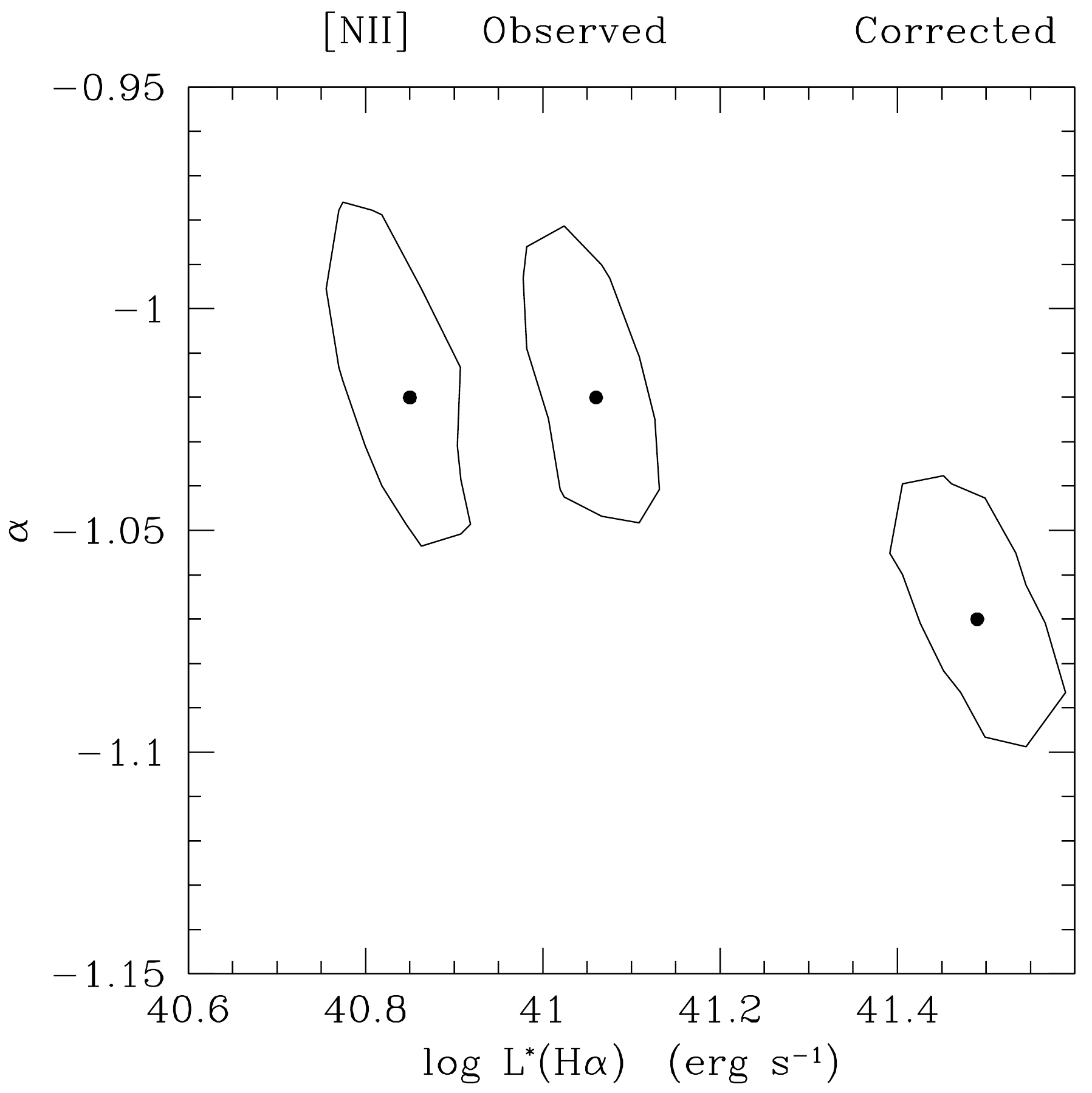}\\
\caption{Observed (upper left), corrected for [NII] contamination (upper right), and corrected for [NII] contamination and dust attenuation (lower left) H$\alpha$ 
luminosity functions of the Virgo cluster. Filled black dots show the whole Virgo cluster sample, and filled red and blue dots
show the HI-deficient ($HI-def$ $>$ 0.4) and HI-normal ($HI-def$ $\leq$ 0.4) cluster galaxies. 
The black and red dots overlap for $L(H\alpha)$ $<$ 10$^{38}$ erg s$^{-1}$.
The vertical dot-dashed black lines indicate the expected H$\alpha$ luminosity for a single O3 and B0 star, respectively.  
The solid black line and dotted grey lines show the best-fit and 1$\sigma$ confidence regions for the Schechter luminosity function parametrisation.
The small panels in the top right corner indicate the 1$\sigma$ probability distribution of the fitted Schecheter function parameters in the three luminosity
functions. The same parameters are compared in the lower right panel.}
\label{LFHa}%
\end{figure*}

\subsection{Dust attenuation}

The dust attenuation was estimated using two independent techniques, as extensively described in Boselli et al. (2023).
The first technique relies on the determination of the Balmer decrement, which is possible through the available spectroscopic data described 
in the previous section. For this purpose, we used the same priority as we adopted for the [NII] contamination correction in the choice of
the spectroscopic data. For galaxies without spectroscopic data, we derived the Balmer decrement using the standard scaling relation
between $A(H\alpha)_{BD}$ and the galaxy stellar mass given in Boselli et al. (2023).

The dust attenuation was also derived using the prescription of Calzetti et al. (2010) based on the Wide-field Infrared Survey Explorer 
(WISE; Wright et al. 2010) 22~$\mu$m emission available for all the massive galaxies of the sample.
Finally, to reduce systematic effects in the data, the dust attenuation in the H$\alpha$ line was derived
as

\begin{equation}
{A(H\alpha) = \frac{A(H\alpha)_{BD} + A(H\alpha)_{22\mu m}}{2}}
.\end{equation}

\noindent
The dust attenuation in the H$\alpha$ band is $A(H\alpha)$ $\lesssim$ 3 mag, with most objects having $A(H\alpha)$ $\lesssim$ 0.4 mag,
as shown in Fig. (2) of Boselli et al. (2023).

\subsection{H$\alpha$ luminosities and SFRs}

The H$\alpha$ fluxes corrected for [NII] contamination and dust attenuation were used to measure the 
H$\alpha$ luminosities as described in Boselli et al. (2023). These were derived assuming galaxies at the distance of their associated subcluster structure, as 
defined in Boselli et al. (2014): 16.5 Mpc for cluster A, cluster C, and the low-velocity cloud (LVC), 23 Mpc for cluster B and for the 
W\arcmin\ cloud, and 32 Mpc for the W and M clouds (see Gavazzi et al. 1999; Mei et al. 2007). We rcall that 
Boselli et al. (2014) assumed that the distances of clusters A and C and of the LVC were at 17 Mpc. For consistency 
with other NGVS works, we adopt here 16.5 Mpc for these substructures.

Finally, H$\alpha$ luminosities were converted into SFR using the relation of Calzetti et al. (2010) derived for a Chabrier initial mass function (IMF),

\begin{equation}
{SFR  ~~ [\rm{M_{\odot} yr^{-1}}] = 5.01 \times 10^{-42}  \it{L(H\alpha)} ~~[\rm{erg~s^{-1}}]} 
.\end{equation}

\noindent
We recall that this conversion is valid only whenever several conditions are satisfied:
1) The escape fraction of ionising photons and the number of ionising photons absorbed by dust before ionising the surrounding gas are null
(e.g. Boselli et al. 2009); 2) The contribution to the ionising radiation due to AGNs or evolved stars is negligible;
3) The star formation activity of the galaxy is almost constant on timescales longer than or comparable to that of the typical age 
of the ionising stars ($\gtrsim$ 10 Myr) (stationarity condition, e.g. Boselli et al. 2009);
4) The IMF is sufficiently sampled to avoid stochastic effects (e.g. Fumagalli et al. 2011);
5) The contribution of other heating mechanisms such as thermal conduction and mixing with the surrounding hot IGM,
turbulence due to Kelvin-Helmholtz instabilities, and shock heating created during the stripping process, is negligible.

Point 1) is probably satisfied in fairly normal galaxies such as those observed in the Virgo cluster (values different 
than but still close to zero are generally observed in strong starbursts; Izotov et al. 2016; Leitherer et al. 2016; Chisholm et al. 2018; 
see however Choi et al. 2020, rare in this cluster environment). As extensively discussed in Boselli et al. (2023), condition 2) is 
also satisfied because the number of AGNs in the sample is very limited (four using the BPT diagram; Baldwin et al. 1981; or eight using the
WHAN classification; Cid Fernandes et al. 2011), and in all these objects, the H$\alpha$ emission is largely dominated by the emission of HII regions
in the disc. We also estimated that the contribution of evolved stars to the production of the ionising radiation is only $\simeq$ 2\%\ (Boselli et al. 2023).
Points 3) and 4) are not necessarily satisfied 
in the most extreme dwarf systems detected in the survey. The H$\alpha$ emission in these objects is often dominated 
by one or a few HII regions that might turn on or off , depending on the star formation history of the parent galaxy, 
which at these very low stellar masses ($M_{star}$ $\simeq$ 10$^6$ M$_{\odot}$) is generally very bursty (e.g. Mateo 1998). At the same time, at the faintest
luminosities reached by the VESTIGE survey, which is complete to $L(H\alpha)$ $\geq$ 10$^{36}$ erg s$^{-1}$, we are close to the stochastic 
sampling of the IMF (see  Boselli et al. 2023). We therefore derived the H$\alpha$ and the $SFR$ luminosity 
functions of the cluster simultaneously, keeping in mind that the faint end of the $SFR$ luminosity function ($L(H\alpha)$ $\lesssim$ 10$^{37}$ erg s$^{-1}$
corresponding to $SFR$ $\lesssim$ 5 $\times$ 10$^{-5}$ M$_{\odot}$ yr$^{-1}$) might be affected by these effects.
Finally, the possible contribution of other processes to the ionisation of the gas mentioned in point 5) is hardly quantifiable given the very complex physics
that describe them and the difficulty of including them in tuned models and simulations (e.g. Melekh et al. 2015; Sander \& Hensler 2021). 
The clumpy structure of the H$\alpha$ images similar to that observed in other photometric bands sensitive to the young stellar distribution
(far- (FUV) and near-ultraviolet (NUV), $u$-band) means that their contribution to the total H$\alpha$ emission of the Virgo galaxies analysed in this work 
remains negligible.

We also recall that the limiting luminosity of the VESTIGE survey, $L(H\alpha)$ $\geq$ 10$^{36}$ erg s$^{-1}$, is lower than the H$\alpha$ 
luminosity expected for the ionisation due to a single O star and is rather comparable to that of a single early-B star. These 
are the stars with the lowest mass and temperature that are able to produce ionising photons that cause the H$\alpha$ emission in 
star-forming regions (Sternberg et al. 2003). The H$\alpha$ luminosity of a single star can be derived by converting the photon ionisation rate $Q(H^0)$
into a SFR using eq. 2 of Kennicutt (1998), and then into $L(H\alpha)$ assuming an IMF (Chabrier in this work).
Values of $Q(H^0)$ for different ionising stars are available in Sternberg et al. (2003). For comparison with the H$\alpha$ luminosity function 
of Virgo cluster galaxies, we plot in Figs. 1, 3, and 4 the expected $L(H\alpha)$ for an O3 and B0 class III star.
VESTIGE is able to detect the emission of any single ionising star, and
is thus sampling the whole dynamic range of the H$\alpha$ luminosity function of galaxies in the Virgo cluster.

\section{H$\alpha$ luminosity function}

\subsection{H$\alpha$ luminosity function of the Virgo cluster}

We derived the differential H$\alpha$ luminosity function of the Virgo cluster by counting the number of galaxies per bin of H$\alpha$ luminosity 
(0.5 dex in log scale) detected by VESTIGE. Because of the sensitivity of the survey, which is able to detect the emission of a single 
late-B ionising star, and the applied source identification technique based on optical priors of stellar mass up to $\simeq$ 2 orders of magnitude 
below the expected H$\alpha$ detection limit suggested by the main-sequence relation (Boselli et al. 2023), we considered the VESTIGE survey complete 
down to $L(H\alpha)$ $\simeq$ 10$^{36}$ erg s$^{-1}$. We thus do not apply any completeness correction.

The volume covered by the survey is the one included within a pyramid with a base equal to the VESTIGE footprint (104 deg.$^2$)
and with a height corresponding to the width of the narrow-band filter (3000 km s$^{-1}$, corresponding to 42.9 Mpc for $H_o$ = 70 km s$^{-1}$ Mpc$^{-1}$).
This leads to a volume of 851.7 Mpc$^3$. This volume is significantly larger than the volume of the main body of the Virgo cluster,
$r_{200}$ = 1.55 Mpc (Ferrarese et al. 2012). For consistency with previous NGVS and VESTIGE works we assume throughout this analysis
$r_{200}$ = 1.55 Mpc; this value is slightly larger then the one recently presented in Boselli et al. (2022a), $r_{200}$ = 0.974 Mpc,
and should thus be considered only as an upper limit. Because Virgo has a very complex 3D structure that is formed by several subgroups located at different distances (e.g. de Vaucouleurs 1961; Binggeli et al. 1987; Gavazzi et al. 1999; Solanes et al. 2002; Mei et al. 2007; Boselli et al. 2014b), we cannot accurately estimate its
volume. For this reason, the luminosity functions derived in this work and their parametric fits were derived using the total number of objects without any
normalisation.

Figure \ref{LFHa} shows the luminosity function of the observed H$\alpha$+[NII] luminosity ($L(H\alpha)_{obs}$), the 
H$\alpha$ luminosity corrected for [NII] contamination ($L(H\alpha)_{[NII]}$), and that corrected for [NII] contamination and dust attenuation 
($L(H\alpha)_{cor}$). These H$\alpha$ densities were derived within bins of 0.5 dex using the full sample of H$\alpha$ detected
galaxies (384 sources). We derived the luminosity function for the whole sample and separately for the HI-normal ($HI-def$ $\leq$ 0.4) and the
HI-deficient ($HI-def$ $>$ 0.4) galaxies, the former representing unperturbed systems similar to those populating the field, the latter 
perturbed objects typical of high-density environments.
We fit the luminosity function using a Schechter (1976) function of the form

\begin{equation}
{\rm{\Phi}(L) = \rm{ln(10)\Phi^*10}^{(L-L^*)(1+\alpha)}\rm{exp}(-10^{(L-L^*)})}
\end{equation}
 
\noindent
following the formalism of Metha et al. (2015) and Fossati et al. (2021), deriving the posterior distribution and the best-fit parameters using the MULTINEST Bayesian algorithm 
(Feroz \& Hobson 2008; Feroz et al. 2019). Because the observed shape of the non-parametric luminosity function abruptly decreases 
below $L(H\alpha)$ $\lesssim$ 10$^{38}$ erg s$^{-1}$, we limited the fit to galaxies with $L(H\alpha)$ $\geq$ 10$^{38}$ erg s$^{-1}$ (340 objects). 
The best fit to the data, derived for the observed and corrected H$\alpha$ luminosity functions, is shown in Fig. \ref{LFHa}, 
and the best-fit parameters are listed in Table \ref{fit}. Figure \ref{LFHa} shows that the H$\alpha$ luminosity function of the cluster steeply decreases at the bright 
end, above $L(H\alpha)$ $\gtrsim$ 10$^{41}$ erg s$^{-1}$,  mildly increases in the range 10$^{38.5}$ $\lesssim$ $L(H\alpha)$ $\lesssim$ 10$^{41}$ erg s$^{-1}$,
and abruptly decreases at lower luminosities. It also shows that the luminosity function is dominated by HI-poor systems. Finally, this result is robust with respect to the 
adopted corrections of the observed flux for [NII] contamination and dust attenuation because these corrections are substantial only in objects with the
highest H$\alpha$ luminosities (massive galaxies), which are not numerous in this sample. The slope of the faint end does not significantly change in the 
three luminosity functions, while $L^*(H\alpha)$ obviously does given the nature of the applied corrections (Fig. \ref{LFHa}).

\begin{table*}
\caption{H$\alpha$ luminosity function in the different cluster subsamples }
\label{fitgroups}
{
\[
\begin{tabular}{ccccc}
\hline
\noalign{\smallskip}
\hline
sample                                  & $\alpha$                      & log$L^*(H\alpha)$               & log$\phi^*$   & N.obj \\
\hline  
Virgo                                   & -1.07$^{-1.04}_{-1.10}$       & 41.49$^{41.60}_{41.40}$ & 1.54$^{1.61}_{1.47}$& 340     \\
$R\leq$ 0.5$\times$ $r_{200}$           & -1.03$^{-0.97}_{-1.09}$       & 41.61$^{41.85}_{41.42}$ & 0.94$^{1.08}_{0.79}$& 75      \\
$R>$ 0.5$\times$ $r_{200}$              & -1.08$^{-1.04}_{-1.12}$       & 41.45$^{41.58}_{41.34}$ & 1.42$^{1.51}_{1.33}$& 265     \\
Virgo $HI-def\leq0.4$                   & -0.88$^{-0.80}_{-0.96}$       & 41.60$^{41.80}_{41.42}$ & 1.05$^{1.17}_{0.91}$& 53      \\
Virgo $HI-def>0.4$                      & -1.09$^{-1.06}_{-1.13}$       & 41.40$^{41.52}_{41.30}$ & 1.44$^{1.52}_{1.36}$& 287     \\
Cluster A + B                           & -1.07$^{-1.04}_{-1.11}$       & 41.55$^{41.67}_{41.44}$ & 1.42$^{1.50}_{1.33}$& 266     \\
Cluster C, W + W$^{\prime}$ clouds      & -0.85$^{-0.71}_{-0.98}$       & 40.94$^{41.21}_{40.72}$ & 0.95$^{1.11}_{0.75}$& 33      \\
M + LVC clouds                          & -1.25$^{-1.16}_{-1.34}$       & 41.76$^{42.18}_{41.40}$ & 0.22$^{0.48}_{-0.07}$& 41     \\
\noalign{\smallskip}
\hline
\end{tabular}
\]
Notes: H$\alpha$ luminosities are corrected for [NII] contamination and dust attenuation. 
Units are erg s$^{-1}$ for the H$\alpha$ luminosities. 
}
\end{table*}

\subsection{H$\alpha$ luminosity function of the different cluster substructures}

\begin{figure}
\centering
\includegraphics[width=0.5\textwidth]{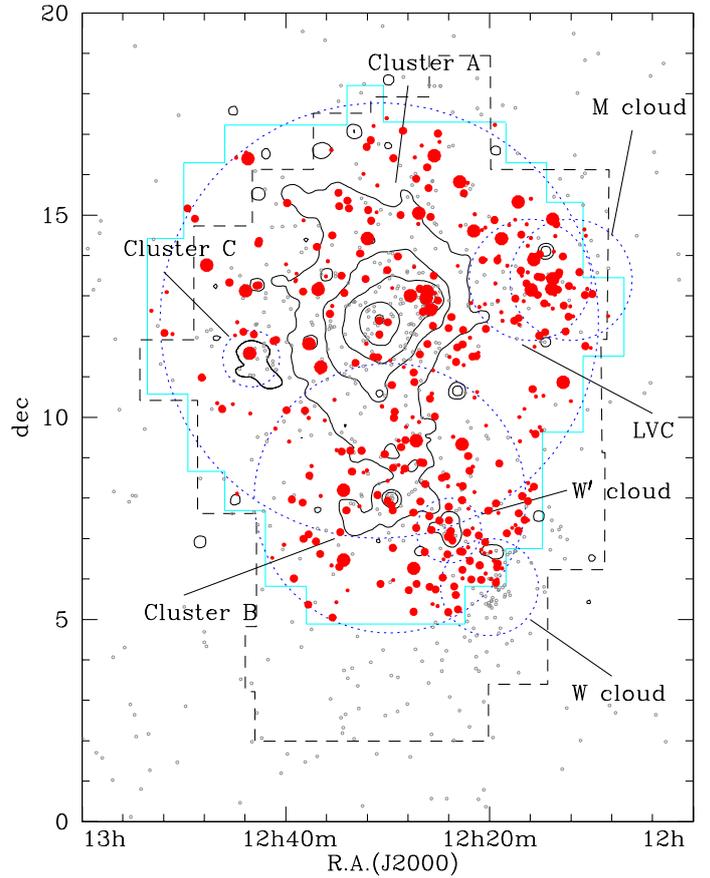}
\caption{Sky distribution of the H$\alpha$ emitting galaxies with recessional velocity $v_{hel}$ $\leq$ 3000 km s$^{-1}$ (filled red dots) and of the 
UV selected galaxies at the distance of the Virgo cluster, these last taken from Boselli et al. (2014). The size of the filled red dots is 
proportional to the H$\alpha$ luminosity of the VESTIGE-detected galaxies: Large symbols show galaxies with $L(H\alpha)$ $>$ 10$^{41}$ erg s$^{-1}$,
medium symbols show objects with 10$^{39}$ $<$ $L(H\alpha)$ $\leq$ 10$^{41}$ erg s$^{-1}$, and small symbols shows galaxies with 
$L(H\alpha)$ $\leq$ 10$^{39}$ erg s$^{-1}$. The solid cyan contour shows the VESTIGE footprint, the dashed black contour the VCC footprint, and the
solid black contours show the X-ray diffuse emission of the cluster from B\"ohringer et al. (1994). The different cluster substructures given in Table \ref{groups}
are indicated with the short-dashed blue circles.
}
\label{angdistLFHa}%
\end{figure}

\begin{figure*}
\centering
\includegraphics[width=0.49\textwidth]{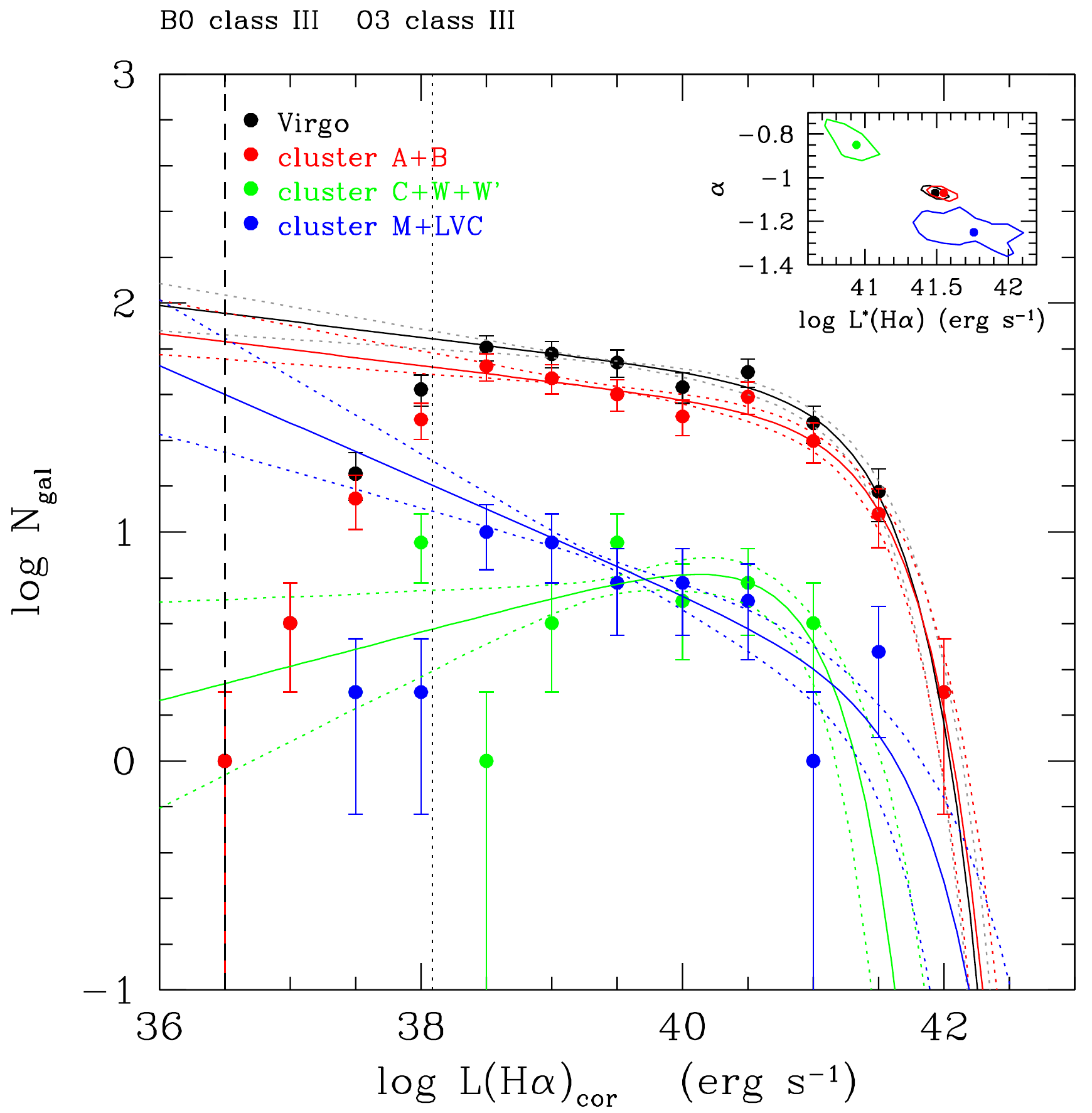}
\includegraphics[width=0.49\textwidth]{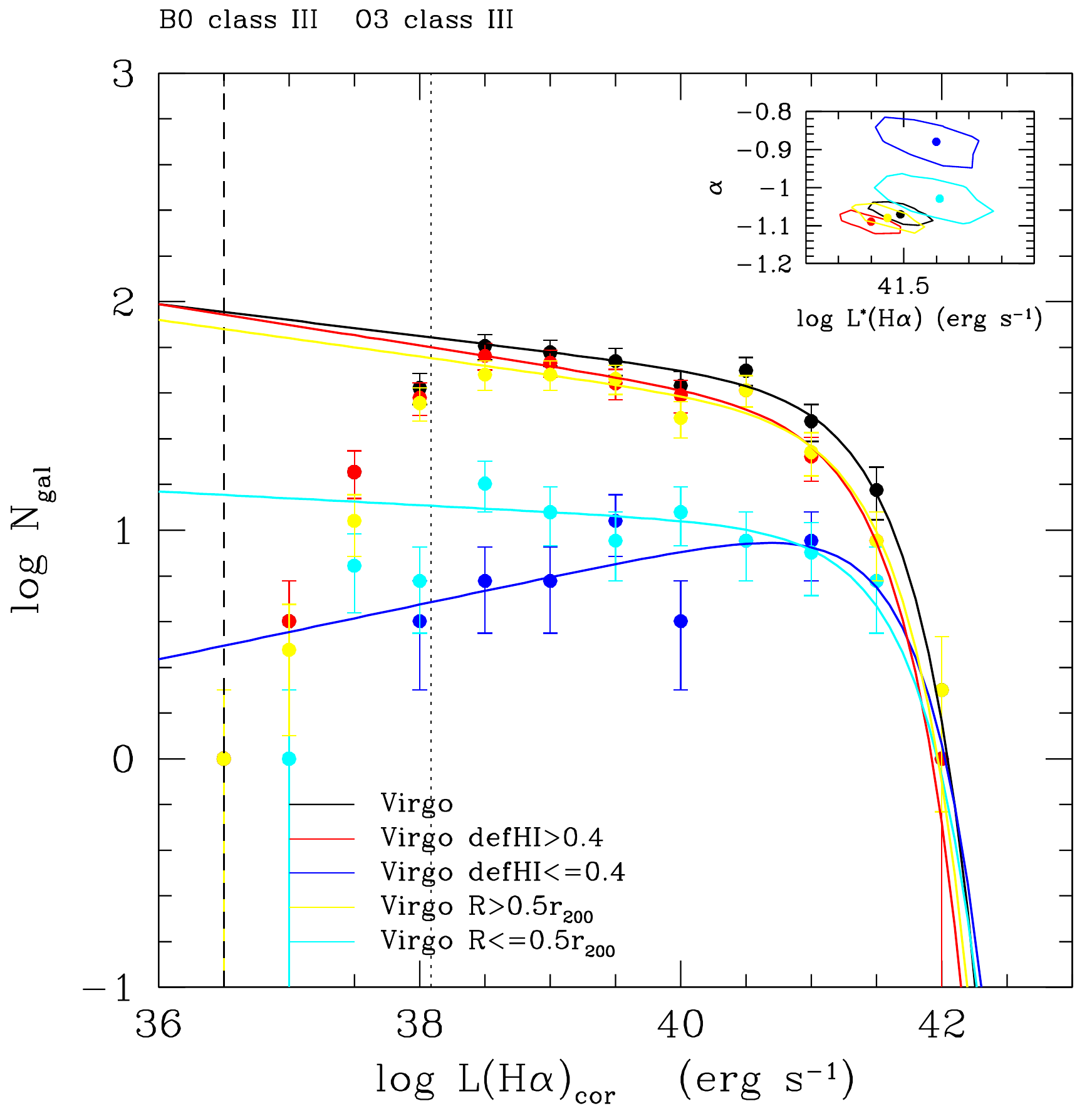}\\
\caption{H$\alpha$ luminosity functions of the Virgo cluster: Left panel: H$\alpha$ luminosity function corrected for [NII] contamination and dust attenuation 
and its Schechter parametrisation (solid and dotted lines, representing the 1$\sigma$ confidence)
for the whole member sample (black) compared to that derived for clusters A and B (red), for cluster C, W, and W$^{\prime}$ clouds (green), 
and for the M and LVC clouds (blue). Black and red dots overlap for $L(H\alpha)$ $\leq$ 10$^{37}$ and $L(H\alpha)$ = 10$^{42}$
erg s$^{-1}$. Right panel: Same H$\alpha$ luminosity function for the whole member sample (black) 
compared to that derived for gas-rich ($HI-def$ $\leq$ 0.4, blue) and gas-poor ($HI-def$ $>$ 0.4, red) galaxies and for objects located within (cyan)
and outside (yellow) $R$ = 0.5$\times$ $r_{200}$, where $R$ is the distance from M87, the centre of cluster A. Black, red, and yellow
dots overlap at some luminosities.
The small panels in the top right corner indicate the 1$\sigma$ probability distribution of the fitted Schecheter function parameters.
The vertical dotted and dashed black lines indicate the expected H$\alpha$ luminosity for a single O3 and B0 star, respectively.  }
\label{LFHagroups}%
\end{figure*}

Figure \ref{LFHagroups} shows the H$\alpha$ luminosity function corrected for [NII] contamination and dust attenuation 
of the whole cluster and compares it to that derived for the two main substructures
of Virgo (clusters A and B) and to other minor substructures (clusters C, W, W$^{\prime}$, M, and the LVC). All these substructures 
are defined as in Boselli et al. (2014) and are shown in Fig. \ref{angdistLFHa}. Membership to these substructures is defined as indicated in Table \ref{groups}.

\begin{table*}
\caption{Properties of the different cluster substructures (see Fig. \ref{angdistLFHa}).}
\label{groups}
{
\[
\begin{tabular}{ccccccccc}
\hline
\noalign{\smallskip}
\hline
Substructure    & R.A.(J2000)   & dec   & Radius$^a$    & velocity range$^b$    & Dist$^c$        & $<vel>$       & $\sigma$       &Central galaxy \\
                & $^o$          & $^o$  & $^o$          & km s$^{-1}$           &Mpc            & km s$^{-1}$     &  km s$^{-1}$   &   \\
\hline
Cluster A       & 187.71        & 12.39 & $<$5.383      & $<$ 3000              & 16.5            & 955           & 799            & M87   \\
Cluster B       & 187.44        & 8.00  & $<$3.334      & $<$ 3000              & 23              & 1134          & 464            & M49   \\
Cluster C       & 190.85        & 11.45 & $<$0.7        & $<$ 3000              & 16.5            & 1073          & 545            & M60   \\
W cloud         & 185.00        & 5.80  & $<$1.2        &1000 $< v_{hel} <$ 3000& 32             & 2176          & 416            & NGC4261       \\
W' cloud        & 186.00        & 7.20  & $<$0.8        & $<$ 2000              & 23              & 1019          & 416            & NGC4365       \\
M cloud         & 183.00        & 13.40 & $<$1.5        &1500 $< v_{hel} <$ 3000& 32             & 2109          & 280            & NGC4168       \\
LVC cloud       & 184.00        & 13.40 & $<$1.5        & $v_{hel}$ $<$400      & 16.5            & 85            & 208            & NGC4216       \\
\noalign{\smallskip}
\hline
\end{tabular}
\]
Note: Galaxies in the overlapping regions that satisfy the membership criteria of two different structures are assumed to be members of the smallest structure
for clusters C, W, W', M, and the LVC clouds. Those in the overlapping region between clusters A and B are assigned to the substructure whose
centre is closest to the galaxy.\\ 
$a$: Radius used to identify the different substructures. For clusters A and B, this corresponds to the $r_{200}$ radius from Ferrarese et al. (2012).
$b$: Velocity range up to 3500 km s$^{-1}$ taken in Boselli et al. (2014) to include field galaxies at the periphery of the cluster. 
Only one object within the VESTIGE footprint with 3000 $<$ $v_{hel}$ $<$ 3500 km s$^{-1}$ is clearly detected by VESTIGE, 
VCC 357 ($v_{hel}$ = 3008 km s$^{-1}$), here considered as a Virgo cluster member.
$c$: Mean distance of each single substructure.}
\end{table*}

To increase the statistics in the different clouds, we separated the galaxies belonging to clusters C, W, and W$^{\prime}$ clouds
from those belonging to the M and LVC clouds because the first subsample is dominated by early-type systems and gas-deficient spirals and 
the second one by gas-rich star-forming objects (Gavazzi et al. 1999). The best-fit parameters 
again measured for galaxies with $L(H\alpha)$ $\geq$ 10$^{38}$ erg s$^{-1}$ 
are given in Table \ref{fitgroups}. Despite the limited statistics in
clusters C, W, and W$^{\prime}$ clouds (33 objects) and in the M and LVC clouds (41 galaxies), Fig. \ref{LFHagroups} suggests that the slope  
of the H$\alpha$ luminosity function in the luminosity range 10$^{38}$ $<$ $L(H\alpha)$ $\leq$ 10$^{40}$ erg s$^{-1}$
is steeper in the M and LVC clouds ($\alpha$ = -1.25$^{-1.16}_{-1.34}$), which are dominated by gas-rich star-forming galaxies, than in the more relaxed clusters
A and B ($\alpha$ = -1.07$^{-1.04}_{-1.11}$) or in the other smaller substructures (clusters C, W, and W' clouds, $\alpha$ = -0.85$^{-0.71}_{-0.98}$)
(the marginalised PDFs of $\alpha$ in the M+LVC and C+W+W$^{\prime}$ subsamples are statistically different at $\simeq$ 4$\sigma$ significance). 
Figure \ref{LFHagroups} also shows the luminosity function derived for gas-rich ($HI-def$ $\leq$ 0.4)
and gas-poor ($HI-def$ $>$ 0.4) galaxies and shows them separately for objects located within 
and outside $R$ = 0.5$\times$$r_{200}$, where $R$ is the distance from M87, the centre of cluster A. Clearly, star-forming galaxies are lacking 
principally in the inner regions of the cluster. The gas-rich star-forming systems are also lacking, principally those with low H$\alpha$ luminosities.
These results do not depend on the reduced sensitivity of the VESTIGE survey in the outskirts of the cluster, where completeness is 
reached for $L(H\alpha)$ $\geq$ 2 $\times$ 10$^{36}$ erg s$^{-1}$.

\subsection{Star formation rate function and comparison with other tracers}

\begin{figure}
\centering
\includegraphics[width=0.5\textwidth]{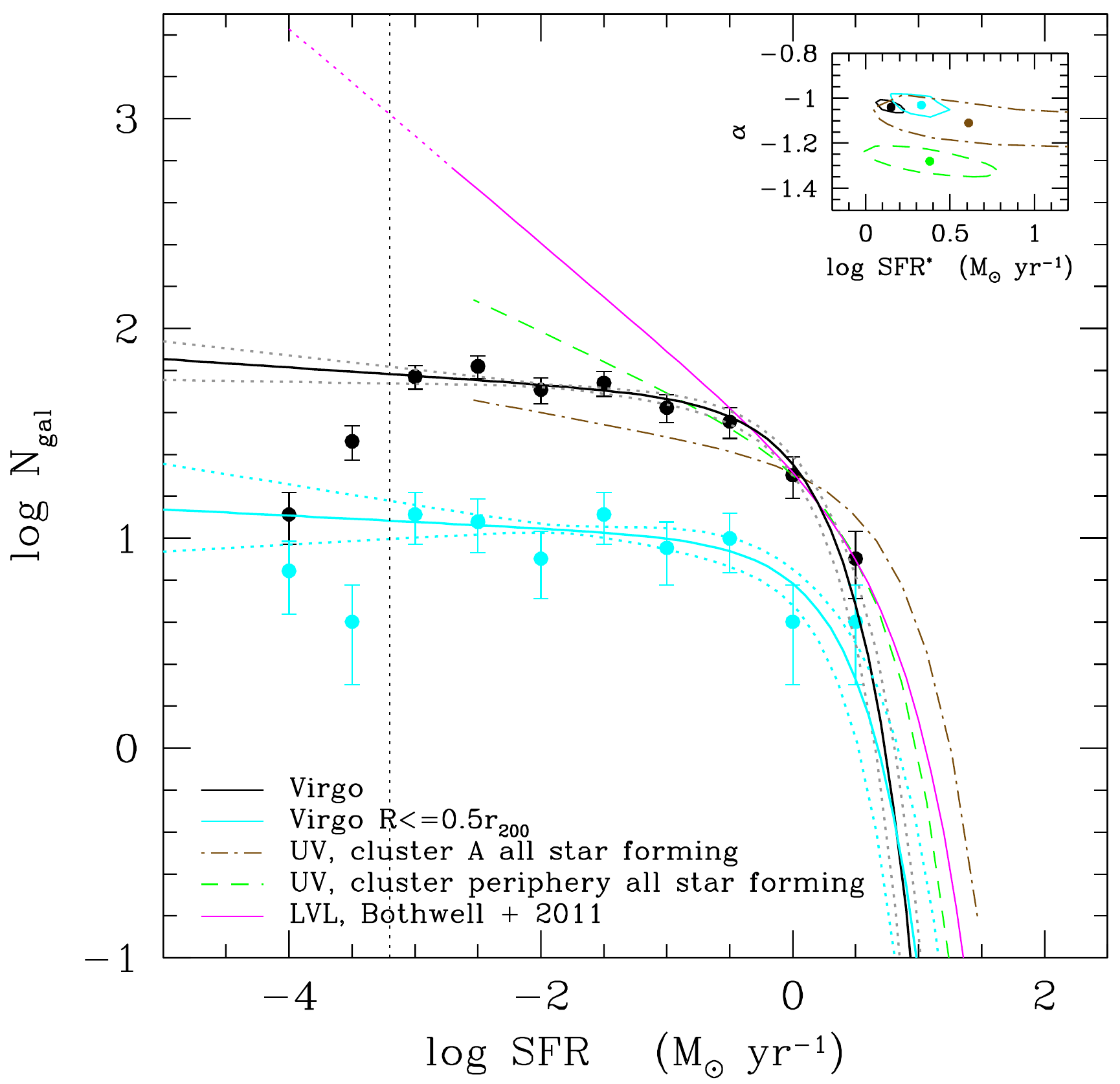}
\caption{Comparison of the $SFR$ luminosity functions of the Virgo cluster derived using the VESTIGE H$\alpha$ 
imaging data (filled dots, solid and dotted lines, representing the 1$\sigma$ confidence interval, in black
for the whole cluster, and in cyan for galaxies within the inner $R$ $\leq$ 0.5$\times r_{200}$) and using GALEX NUV data extracted from the GUViCS survey
(Boselli et al. 2011, 2016c). NUV-derived SFR luminosity
functions are for the cluster periphery (galaxies at a distance from M87 $>$ 1.1$\times r_{200}$, dashed green line)
and for galaxies in the inner cluster A region ($<$ 0.5$r_{200}$, dot-dashed brown line). The solid magenta line 
shows the SFR function derived by Bothwell et al. (2011) for the LVL survey corrected to a Chabrier IMF. 
This function is accurately estimated down to log $SFR$ $\simeq$ -2.7 M$_{\odot}$ yr$^{-1}$, and it is poorly constrained down 
to log $SFR$ $\simeq$ -4 M$_{\odot}$ yr$^{-1}$ (dotted magenta line).
These functions are normalised to the number of galaxies within the Virgo 
cluster sample in the bin centred at $SFR$ = 1 M$_{\odot}$ yr$^{-1}$.
The vertical dotted black line shows the limit in $SFR$ adopted for the fit.   
}
\label{LFSFR}%
\end{figure}

Figure \ref{LFSFR} shows the SFR function derived in this work using VESTIGE H$\alpha$ imaging data and compares it
with the one derived using GALEX ultraviolet (UV) data of the GALEX ultraviolet Virgo cluster survey (GUViCS; Boselli et al. 2011). For this exercise, we excluded galaxies M87, NGC 4262, and NGC 4552, in which the H$\alpha$ emission is due to processes other than photoionisation of young massive stars (Boselli et al. 2019, 2022b).
The best-fit parameters of the Schechter function are given
in Table \ref{fitSFR}. Since the GUViCS data are taken in a wider region ($\simeq$ 300 deg.$^2$ 
vs. $\simeq$ 104 deg.$^2$ for VESTIGE), they were used to derive the luminosity function separately for the cluster periphery (at a distance from M87
$>$ 6.1$^{\circ}$, corresponding to $\gtrsim$ 1.1$\times$$r_{200}$)
and for galaxies belonging to the inner regions of the main body of cluster A ($<$ 0.5$\times$$r_{200}$, see Boselli et al. 2014). These
last are comparable to those derived in this work using the H$\alpha$ galaxies located in the inner cluster, marked in cyan in Fig. \ref{LFSFR}. As for the 
previous comparisons, however, the GUViCS data are normalised to the number of objects detected by VESTIGE in the bin $SFR$ = 1 M$_{\odot}$ yr$^{-1}$.

The uncertainties in the derivation of the best-fit parameters of the H$\alpha$ derived star formation function are significantly lower
than those estimated using the GALEX UV data.
Figure \ref{LFSFR} shows that the UV derived SFR  function of the cluster periphery is significantly steeper at the faint end than the 
one derived using H$\alpha$ data ($\alpha_{NUV}$ = -1.28 vs. $\alpha_{H\alpha}$ =  -1.04$^{-1.01}_{-1.07}$). At the bright end, 
the NUV luminosity function for cluster A gives more galaxies, although the overall shape of the NUV luminosity function is still consistent with the observed number of H$\alpha$
emitting sources. The difference at the bright end is more pronounced when the H$\alpha$ derived function is compared to the NUV function derived for cluster A,
which gives more galaxies with a high star formation activity. At the faint end, the two distributions are fairly consistent
($\alpha_{NUV}$ = -1.11 vs. $\alpha_{H\alpha}$ = -1.03$^{-1.10}_{-0.97}$ for galaxies located at $R$ $\leq$ 0.5$\times$$r_{200}$). 
The limited depth of the GUViCS survey does not allow us to reach galaxies with SFRs 
log $SFR$ $\lesssim$ -2.6 M$_{\odot}$ yr$^{-1}$, and thus it cannot confirm the abrupt decrease in activity observed in the H$\alpha$ data below this limit.

The VESTIGE SFR function is also compared to the SFR derived for the Local Volume Legacy (LVL) survey sample by Bothwell et al. (2011)
using a combination of UV, IR, and H$\alpha$ data (bivariate luminosity function; Fig. \ref{LFSFR}). This SFR function is well fitted by a Schechter function with 
$SFR^*$ = 5.8 M$_{\odot}$ yr$^{-1}$ (for a Chabrier IMF) and $\alpha$ = -1.51$\pm$0.08 down to its completeness limit of 
log $SFR$ $\simeq$ -2.7 M$_{\odot}$ yr$^{-1}$, but it still increases with a similar slope down to the detection limit of 
log $SFR$ $\simeq$ -4 M$_{\odot}$ yr$^{-1}$. Clearly, the number of star-forming objects in the field is significantly higher than that in the
Virgo cluster for $SFR$ $\lesssim$ 0.1  M$_{\odot}$ yr$^{-1}$.

\begin{table}
\caption{Best-fit parameters for the SFR luminosity function}
\label{fitSFR}
{
\[
\begin{tabular}{ccccc}
\hline
\noalign{\smallskip}
\hline
Variable                & sample                                & $\alpha$                      & log$SFR$                        & log$\phi^*$   \\
\hline  
$SFR$                   & log$SFR$ $\geq$ -3.2                  & -1.04$^{-1.01}_{-1.07}$       & 0.15$^{0.25}_{0.07}$            & 1.59$^{1.65}_{1.52}$\\
\noalign{\smallskip}
\hline
\end{tabular}
\]
Notes: units are in M$_{\odot}$ yr$^{-1}$ for the SFRs.
}
\end{table}

\section{Discussion}

\subsection{ H$\alpha$ luminosity function of the Virgo cluster}

The exceptional quality of the VESTIGE survey in terms of sensitivity and image quality allowing us to derive the H$\alpha$ luminosity function of the Virgo cluster for the first time. This exercise was made on 
a statistically significant sample of 384 Virgo cluster members. Because of the depth of the survey, 
this sample is complete and includes all possible emitting sources where the gas is 
ionised by massive stars. The unique set of available multifrequency data allowed us to correct the VESTIGE narrow-band imaging data 
for [NII] contamination and dust attenuation with high accuracy. The analysis presented in Sec. 4, and in particular in Fig. \ref{LFHa}, clearly indicates 
that the derivation of the luminosity function is robust compared to these corrections, which only mildly modify the shape of both the 
non-parametric and parametric (Schechter) functions. For this reason, the data also allow us to derive the SFR function of the cluster very well. 

The non-parametric luminosity function is well represented by a Schechter function in the luminosity range
10$^{38}$ $\lesssim$ $L(H\alpha)$ $\lesssim$ 10$^{42}$ erg s$^{-1}$ and does not require other parametrisations that are sometime used
to characterise the field luminosity function (Saunders's formalism; Gunawardhana et al. 2013). The luminosity function
is poorly constrained at the bright end because only a limited volume is sampled (851.7 Mpc$^3$), but it is excellent at the faint end.
The abrupt decrease at $L(H\alpha)$ $\lesssim$ 10$^{38}$ erg s$^{-1}$ is probably real for the following reasons. First of all, 
this limiting luminosity is two orders of magnitude above the detection limit of the survey ($L(H\alpha)$ $\simeq$ 10$^{36}$ erg s$^{-1}$)
derived by measuring the detection rate of fake sources of similar luminosity injected in real images (Boselli et al. 2018a).
Second, H$\alpha$ sources have been searched for in all the stellar continuum-subtracted frames in which galaxies have been 
detected in the deep optical NGVS frames (Ferrarese et al. 2012). According to the recently derived main-sequence relation (Boselli et al. 2023),
the limiting H$\alpha$ luminosity of $L(H\alpha)$ $\simeq$ 10$^{36}$ erg s$^{-1}$ occurs at about two orders of magnitude above 
the stellar mass sensitivity limit of NGVS. It is thus very unlikely that any H$\alpha$ emitting source associated with an optically
identified galaxy has been lost after visual inspection of all the galaxies identified as Virgo cluster members in the NGVS survey (3869 galaxies).
A preliminary statistical analysis of all line emitters automatically identified using SExtractor used on $\simeq$ 10\%\ of the mapped 
region indicates only one potential Virgo cluster member candidate (see Sec. 2.2).
Possible sources might have been missed whenever their surface brightness in H$\alpha$ is below the detection limit of the survey
($\Sigma(H\alpha)$ $\simeq$ 2 $\times$ 10$^{-18}$ erg s$^{-1}$ cm$^{-2}$ arcsec$^{-2}$) throughout the optical disc of the galaxy. 
This is a very low surface brightness that can be 
reached and overtaken only with a few facilities after very long exposure times (e.g. MUSE at the VLT). It is difficult to quantify the possible 
frequency of galaxies with an H$\alpha$ surface brightness below this threshold.  

\begin{figure}
\centering
\includegraphics[width=0.5\textwidth]{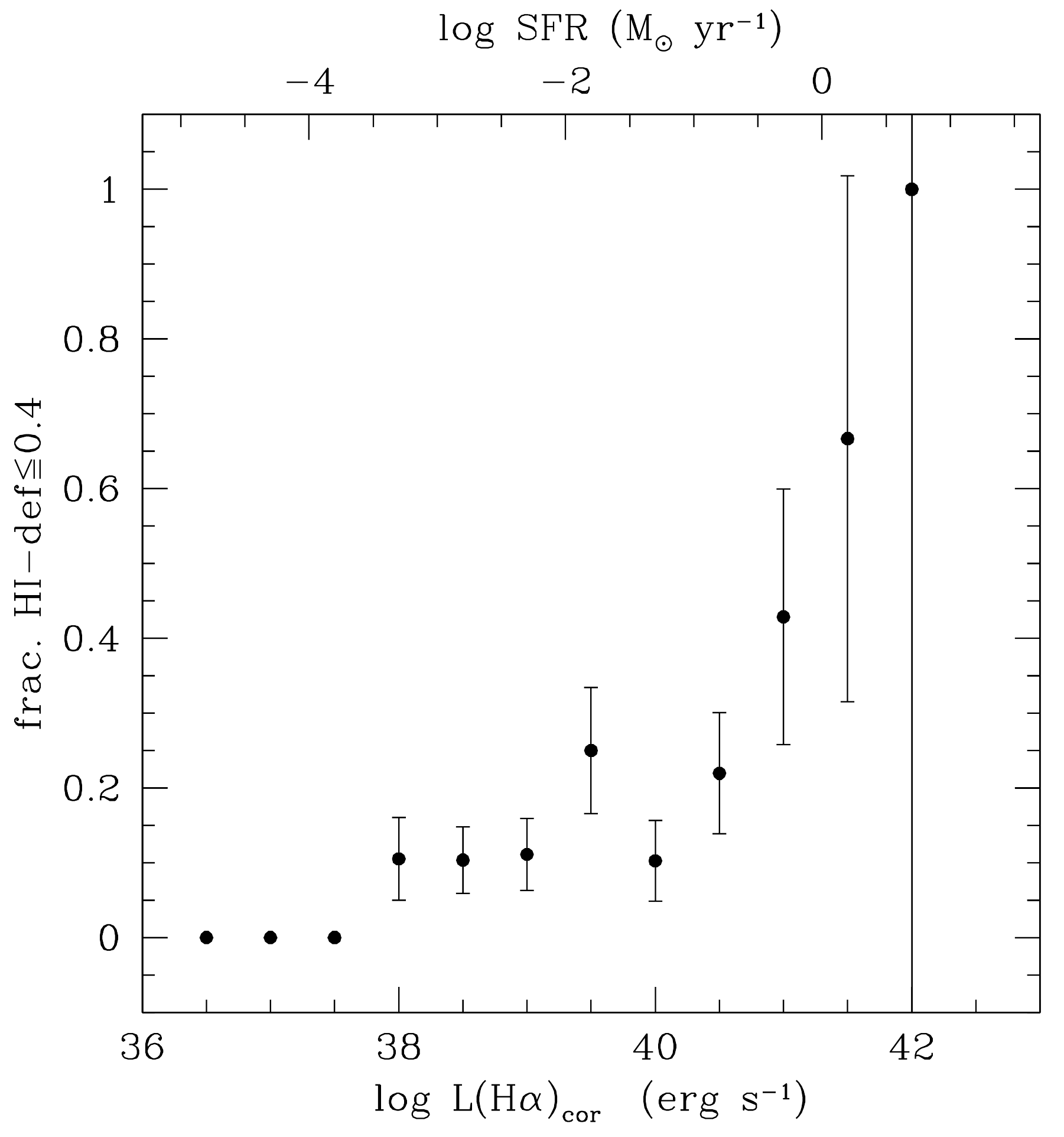}
\caption{Relation of the fraction of Virgo cluster gas-rich galaxies and their corrected H$\alpha$ luminosity (lower X-axis) or SFR (upper X-axis).
}
\label{ratioHI}%
\end{figure}

Figures \ref{LFHa} and \ref{LFHagroups} clearly indicate that the H$\alpha$ luminosity function of Virgo is dominated by gas-poor ($HI-def$ $>$ 0.4) galaxies.
The ratio of gas-rich to gas-poor objects strongly decreases with decreasing H$\alpha$ luminosity, as illustrated in Fig. \ref{ratioHI}. At H$\alpha$ luminosities 
of $L(H\alpha)$ $\lesssim$ 10$^{38}$ erg s$^{-1}$ , there are no HI-rich galaxies, while galaxies with a normal HI gas content dominate 
at $L(H\alpha)$ $\gtrsim$ 10$^{41}$ erg s$^{-1}$. A similar trend is also observed for galaxies located at $R$ $>$ 0.5$\times r_{200}$,
while the large uncertainties due to the poor statistics prevent us from detecting any trend in the inner cluster ($R$ $\leq$ 0.5$\times r_{200}$). 
Figures \ref{angdistLFHa} and \ref{LFHagroups} also show that star-forming galaxies, and in 
particular, low-mass systems, are less frequent in the inner cluster regions. The difference between the SFR function of Virgo, which is
flat at the faint end ($\alpha$ = -1.04$^{-1.00}_{-1.07}$), and of the local volume, which steeply rises with a slope of $\alpha$ = -1.51$\pm$0.08
in the same range of star formation activity (10$^{-4}$ $\leq$ $SFR$ $\leq$ 10$^{-1}$ M$_{\odot}$ yr$^{-1}$) is also clear.

All this observational evidence can be explained considering that external perturbations affecting galaxy evolution in clusters, such as ram-pressure 
stripping (Gunn \& Gott 1972), act principally on the HI content, which constitutes the main gas reservoir feeding star formation.
These stripping mechanisms
are increasingly efficient with the decrease in stellar mass of galaxies (in unperturbed systems traced by the H$\alpha$ luminosity, main-sequence relation) 
because their shallower gravitational potential keeps the gas well anchored to the stellar disc (Boselli et al. 2022a). In the lowest
H$\alpha$ luminosity regime ($L(H\alpha)$ $\lesssim$ 10$^{38}$ erg s$^{-1}$, corresponding to $M_{star}$ $\lesssim$ 10$^7$ M$_{\odot}$), 
the stripping of the gas is total so that the star formation activity is fully quenched on very short timescales (e.g. Boselli et al. 2008, 2014, 2022b, 2023; Junais et al. 2022).
Thus, the probability of observing Virgo cluster galaxies with these very low H$\alpha$  luminosities is very low, while this is not the case in the field (Bothwell et al. 2011).
Despite the poor statistics, Fig. \ref{LFHagroups} also seems to indicate that in the luminosity range 10$^{38}$ $\lesssim$ $L(H\alpha)$ $\lesssim$ 10$^{41}$ erg s$^{-1}$
, the H$\alpha$ luminosity function is almost flat inside all
the relaxed substructures of the cluster that are dominated by early-type systems and that are well identified in the plane of the sky versus redshift distribution (see Boselli et al. 2014).
The slope, in contrast, seems to be significantly steeper (at the 4$\sigma$ level) in the two main infalling regions at the north-west side of the cluster (M and LVC clouds). These clouds are
characterised by gas-rich star-forming systems that might be infalling for the first time into the cluster (Gavazzi et al. 1999).

\subsection{Comparison with other photometric bands}

\begin{figure*}
\centering
\includegraphics[width=0.49\textwidth]{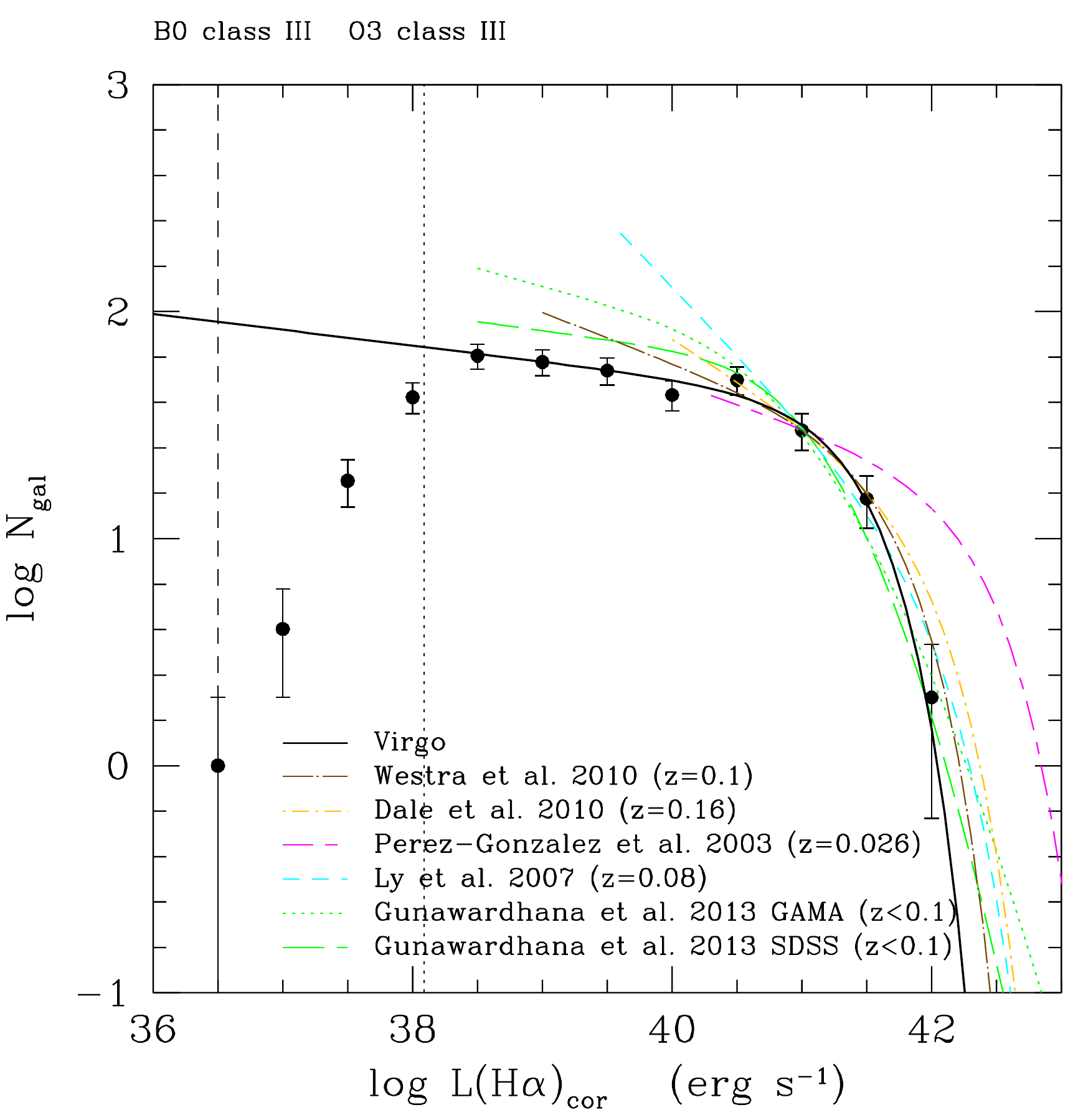}
\includegraphics[width=0.49\textwidth]{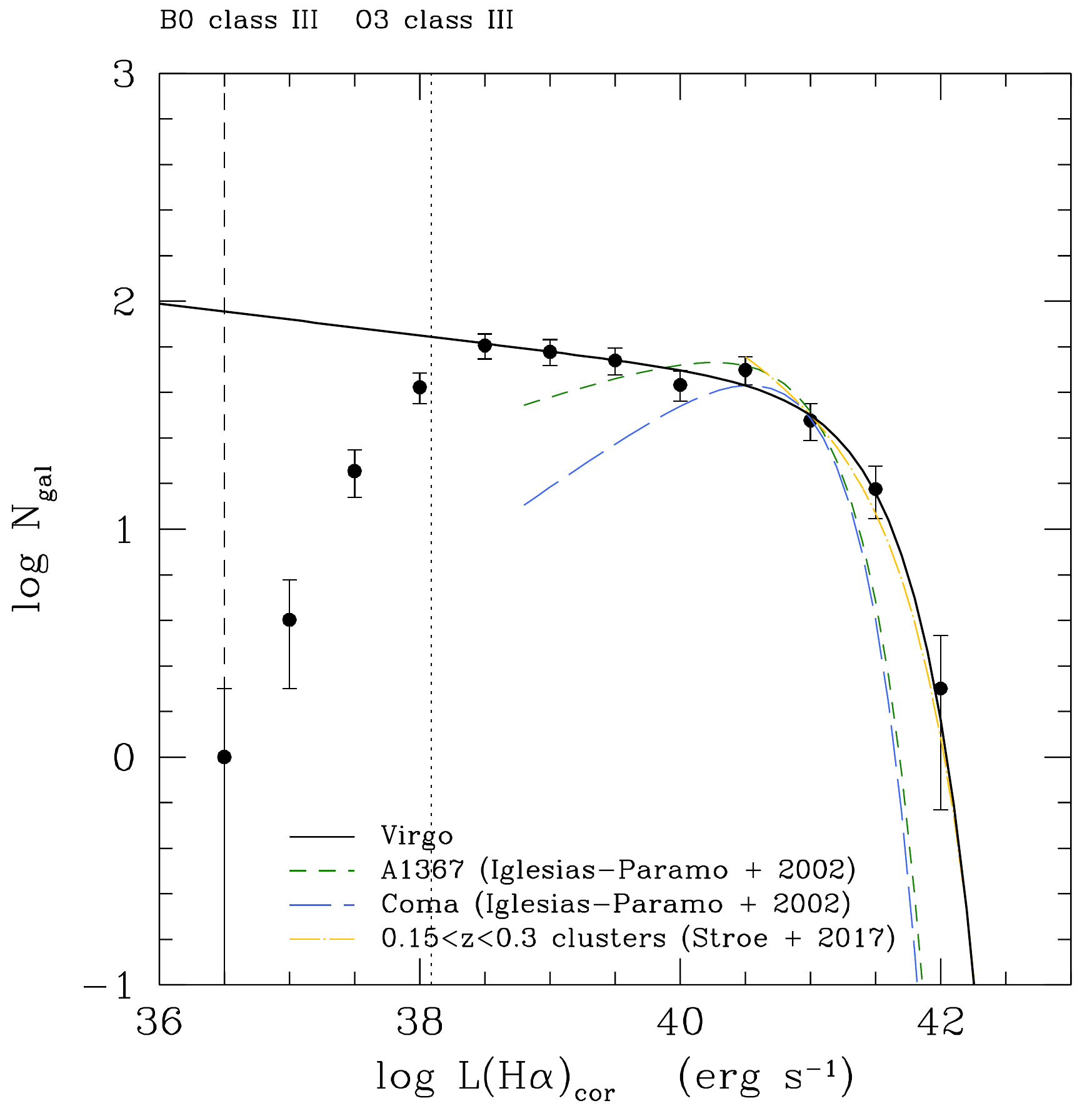}\\
\caption{H$\alpha$ luminosity function of the Virgo cluster corrected for [NII] contamination and dust attenuation (filled black dots)
and its Schechter parametrisation (solid black lines) compared to those derived for field galaxies in the local Universe ($z$ $\lesssim$ 0.16, dashed curves) (left panel) and
star-forming galaxies in nearby clusters (Coma and A1367, from Iglesias-Paramo et al. (2002) and a sample of 
clusters at 0.15$<$ $z$ $<$ 0.3 by Stroe et al. (2017)) (right panel). 
All luminosity functions are normalised to the number of galaxies within the Virgo 
cluster sample in the bin centred at $L(H\alpha)_{cor}$ = 10$^{41}$ erg s$^{-1}$.
The vertical dotted and dashed black lines indicate the expected H$\alpha$ luminosity for a single O3 and B0 star, respectively.  }
\label{LFHafieldcluster}%
\end{figure*}

The H$\alpha$ luminosity function derived in this work can be used quantify the impact of the environment on the star formation process (and its quenching) on strong statistical grounds. 
This can be achieved by comparing the H$\alpha$ luminosity function of Virgo to those available in the literature in other photometric bands, for 
other cluster galaxies, and for the field. This exercise is shown in Fig. \ref{LFHafieldcluster}, but it is clearly restricted
by the lack of data sampling the faint end of the luminosity functions in other nearby clusters or in the field, all limited to
$L(H\alpha)$ $\geq$ 10$^{38.5-39}$ erg s$^{-1}$. 


\begin{figure}
\centering
\includegraphics[width=0.5\textwidth]{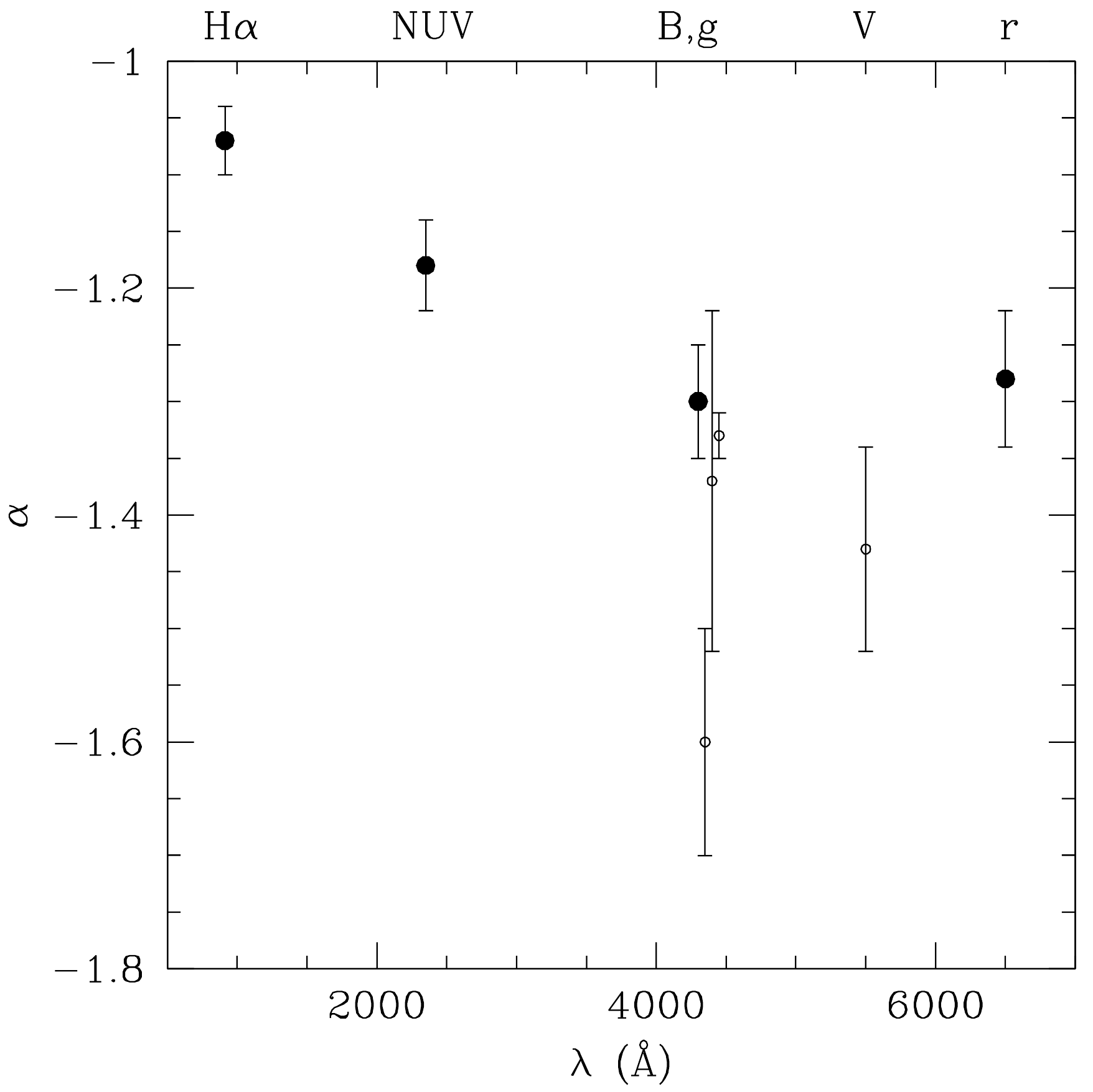}
\caption{Relation of the slope of the faint end $\alpha$ of the luminosity function and the wavelength at which the luminosity function has bean measured.
The large filled dots show values derived within a footprint comparable to the one used in the VESTIGE survey, and small open squares show values measured within a
much more limited region of the cluster. 
The slope of the luminosity function in the different photometric bands was taken from Boselli et al. (2016c) in the UV, Sandage et al. (1985), 
Trentham \& Hodgkin (2002), and Sabatini et al. (2003) in the $B$, Ferrarese et al. (2016) in the $g$, Leider et al. (2012) in the 
$V$, Rines \& Geller (2008) in the $r$. Values in the $B$ and $g$ bands have been slightly shifted in $\lambda$ to avoid confusion.
}
\label{alphalambda}%
\end{figure}

We first compared the slope of the luminosity function derived in this work ($\alpha_{H\alpha}$ = -1.07$^{-1.04}_{-1.10}$) 
with those measured in the Virgo cluster in other bands. The slope derived using the GUViCS data in the NUV, a photometric band 
that is sensitive to the emission of the young stellar population, is $\alpha_{NUV}$ = -1.18 when measured 
within the main body of cluster A ($R$ $\leq$ 0.5$\times r_{200}$), and it is $\alpha_{NUV}$ = -1.22 at the cluster periphery ($R$ $>$ 1.1 $\times r_{200}$;
Boselli et al. 2016c). Steeper slopes are measured on the photographic plate material of the VCC survey within a very similar sky region
($\alpha_{B}$ = -1.30, Sandage et al. 1985), and in the $r$-band when derived using SDSS data for galaxies within 1 Mpc from M87 
($\alpha_{r}$ = -1.28 for $R$ $\simeq$ 0.65 $\times r_{200}$; Rines \& Geller 2008). There are other estimates of the slope, but they are limited to selected regions:
in the very inner 4 deg.$^2$ ($R$ $\leq$ 0.2 $\times r_{200}$), $\alpha_{V}$ $\simeq$ -1.43 (Lieder et al. 2012) and $\alpha_{g}$ = -1.33, but possibly 
increasing to $\alpha_{g}$ = -1.60 when
including ultra-compact dwarf (UCD) galaxies (Ferrarese et al. 2016). Other estimates concern different stripes that extend from the cluster centre to
the periphery (Trentham \& Hodgkin 2002,  $\alpha_{B}$ $\simeq$ -1.37; Sabatini et al. 2003, $\alpha_{B}$ $\simeq$ -1.6). 
All these results suggest a steepening of the 
slope with the ageing of the stellar population within the Virgo cluster, as depicted in Fig. \ref{alphalambda}. In the optical bands, 
the faint end of the luminosity function of Virgo is dominated by early-type quiescent dwarf ellipticals and spheroidals (Sandage et al. 1985;
Ferrarese et al. 2016). For comparison, in the optical bands, the slope of the luminosity function 
in the field is $\alpha$ $\simeq$ -1.0 when it is derived using the SDSS data (Blanton et al. 2003), 
or steeper ($\alpha$ $\simeq$ -1.5) whenever it is corrected to take into account the possible contribution 
of low surface brightness objects (Blanton et al. 2005). In the field, the faint end is dominated by star-forming systems.

\subsection{Comparison with the field and with other clusters}

We compared the properties of the Virgo H$\alpha$ luminosity function with those observed in the field and in other clusters of galaxies (see Fig \ref{LFHafieldcluster}).
The H$\alpha$ luminosity function of field galaxies was derived in the literature using different samples of H$\alpha$ selected or optically selected 
(bivariate luminosity function) objects. These include those derived using the emission line galaxies of the Universidad Complutense de Madrid 
objective-prism survey (UCM; Gallego et al. 1995, 
Perez-Gonzalez et al. 2003), the narrow-band imaging Wyoming Survey for H$\alpha$ (WySH; Dale te al. 2010), 
the Subaru Deep Field (SDF; Fujita et al. 2003; Ly et al. 2007; Morioka et al. 2008),
the HST COSMOS2 field (Shioya et al. 2008), the Calar Alto Deep Imaging Survey (CADIS; Hippelein et al. 2003), the
Taurus Tunable Filter Field Galaxy Survey (Jones \& Bland-Hawthorn 2001),
optically selected galaxies from the Smithsonian Hectospec Lenisng Survey (SHELS; Westra et al. 2010), or from the galaxy and mass assembly survey (GAMA) 
and SDSS (Guanawhardana et al. 2013).

Figure \ref{LFHafieldcluster} (left panel) shows the comparison between the H$\alpha$ luminosity function derived in this work using data corrected 
for [NII] contamination and dust attenuation and other H$\alpha$ luminosity functions for field galaxies in the local Universe 
($z$ $\lesssim$ 0.16). Because of the overdensity of galaxies in the Virgo cluster, a direct comparison of the $\Phi$ parameter 
would be dominated by the very different number density of objects in the two environments and would offer little additional information. 
For a fair comparison of the shapes, all the field Schechter functions were normalised to 
the number of galaxies the Virgo cluster sample has in the bin centred at $L(H\alpha)_{cor}$ = 10$^{41}$ erg s$^{-1}$.
Because of the limited number of bright objects detected in this small sampled volume, we limited the comparison to the slope and characteristic luminosity.
The typical $\alpha$ and $L^*$ parameters of these field H$\alpha$ luminosity functions 
in the local Universe range within -1.59 $\leq$ $\alpha$ $\leq$ -1.20 and 41.74 $\leq$ log$L(H\alpha)^*$ $\leq$ 42.43 erg s$^{-1}$. 
These numbers are significantly (5 and 3 $\sigma$, respectively) different than those derived for the Virgo cluster (see Table \ref{fit}),
with a steeper slope and a brighter characteristic luminosity in field objects with respect to the Virgo cluster galaxies.

Deep wide-field H$\alpha$ narrow-band imaging surveys of clusters are optimal for identifying all emitting sources using an unbiased 
selection. They also have the advantage of providing integrated fluxes in extended sources, which limits any possible systematic effect
in aperture corrections that are substantial in nearby extended sources. Their main limitation, however, is the need of tuned wide-field 
narrow-band expensive filters with a transmissivity curve centred on 
the H$\alpha$ emitting line at the redshift of the targeted clusters. For this reason, the number of nearby clusters observed 
using untargeted narrow-band imaging surveys is very limited. A dozen clusters at redshift $z$ $\lesssim$ 0.3 have been observed so far 
using this technique (e.g. Balogh \& Morris 2000; Couch et al. 2001; Balogh et al. 2002; Umada et al. 2004; Stroe et al. 2017;
see also Liu et al. 2021 for Fourier transform spectrograph imaging), 
while those in the very nearby universe are only Virgo (Boselli et al. 2018a), Coma (Iglesias-Paramo et al. 2002; Yagi et al. 2010), 
and A1367 (Iglesias-Paramo et al. 2002; Yagi et al. 2017). Although deep Subaru narrow-band imaging data are available for both Coma
and A1367 (Yagi et al. 2010, 2017), the difficulty of optimising the subtraction of the stellar continuum emission prevented the determination
of accurate H$\alpha$ fluxes for all the detected cluster members. 

Their fluxes, however, have been measured using shallower narrow-band imaging data gathered with the wide-field camera at the Isaac Newton telescope (INT)  
(Iglesias-Paramo et al. 2002). A fair comparison with these two clusters would require similar sets of data taken at similar cluster-centric distances and at comparable redshift. 
A tentative comparison with Coma and A1367, whose data were taken with narrow-band imaging, is possible, but should be made with care to properly
consider possible systematic biases in the results.
The filter used at the INT telescope is the [SII] rest-frame narrow-band filter, which only partially overlaps with the 
velocity distribution of the galaxies in Coma and A1367. The luminosity functions derived for these two clusters by Iglesias-Paramo et al. (2002)
were calculated after adopting a quite uncertain statistical correction necessary to take this effect into account. Furthermore, 
both clusters were observed on a contiguous 1 deg.$^2$ field, corresponding to $R$ $\lesssim$ 0.5 $\times r_{200}$ and  
$R$ $\lesssim$ 0.6 $\times r_{200}$, respectively. Because the efficiency of most of the perturbing mechanisms depends on the cluster-centric distance, 
this might introduce systematic effects in the comparison. Finally, the H$\alpha$ luminosity function in Virgo is significantly deeper 
($L(H\alpha)$ $\geq$ 10$^{36}$ erg s$^{-1}$) than in Coma and A1367 ($L(H\alpha)$ $\geq$ 10$^{38.5-39}$ erg s$^{-1}$).

Figure \ref{LFHafieldcluster} (right panel) shows the comparison of the Virgo H$\alpha$ luminosity function derived using the VESTIGE data with those
published in Iglesias-Paramo et al. (2002) for Coma and A1367. Because these surveys are limited to the inner cluster regions, where the 
density of galaxies is higher than at the cluster periphery, we normalised the parametric Schechter luminosity functions of Iglesias-Paramo et al. 
(2002) to the Virgo luminosity function by imposing the same number of objects in the bin centred at $L(H\alpha)_{cor}$ = 10$^{41}$ erg s$^{-1}$.
The best-fit parameters of the Schchter function derived by Iglesias-Paramo et al. (2002) scaled to $H_o$ = 70 km s$^{-1}$ Mpc$^{-1}$ are 
$\alpha$ = -0.82 and log $L(H\alpha)^*$ = 41.01 erg s$^{-1}$ within $R$ $\lesssim$ 0.6 $\times r_{200}$ for A1367 and  
$\alpha$ = -0.60 and log $L(H\alpha)^*$ = 40.94 erg s$^{-1}$ within $R$ $\lesssim$ 0.5 $\times r_{200}$ for Coma.
The values of Virgo are intermediate between those derived in these massive clusters and in the field.  
Interestingly, the slope in the optical bands is very steep in Coma, steeper than in the field (de Propris et al. 1998; Mobasher \& Trentham 1998; 
Andreon \& Pello 2000; Balogh et al. 2001; see for a comparison Boselli \& Gavazzi 2006). In  A1367, the slope is $\alpha_{r}$ = -1.07$_{-0.16}^{+0.20}$
(Iglesias-Paramo et al. 2003), comparable to the field but steeper than the one measured in H$\alpha$.

We also compare in Fig. \ref{LFHafieldcluster} the H$\alpha$ luminosity function of Virgo derived in this work with the
composite luminosity function of cluster galaxies at redshfit 0.15 $<$ $z$ $<$ 0.30 derived by Stroe et al. (2017) using
a similar narrow-band imaging technique to identify the emitting sources and measure their total flux.
Figure \ref{LFHafieldcluster} does not show any systematic difference between the bright end of the H$\alpha$ luminosity function of Virgo
galaxies and the composite luminosity function of cluster galaxies at redshift 0.15 $<$ $z$ $<$ 0.30. Because the dynamic range covered by this composite luminosity function is limited, the comparison cannot be made for $L(H\alpha)$ $\lesssim$ 10$^{40.5}$ erg s$^{-1}$.
 
Overall, these results are qualitatively consistent with an evolutionary picture in which galaxies enter the cluster as gas-rich systems,
lose their gas content, and quench their star formation activity on relatively short timescales (shorter than the typical crossing 
time of the cluster) to become quiescent objects without H$\alpha$ emission. The VESTIGE luminosity function only traces the distribution 
of star-forming objects in H$\alpha$, while in the other bands, it simultaneously includes star-forming and quiescent systems.
Perturbing processes such as ram-pressure stripping, often invoked as the dominant process in the Virgo cluster (e.g. Vollmer et al. 2001; 
Boselli et al. 2014, 2023), can explain these observed trends. First of all, they remove the gas from the perturbed galaxies
with an efficiency that increases with the decrease in the total mass of the perturbed object just because dwarf systems have a shallower
gravitational potential well than massive objects that are unable to retain the cold gas anchored to the stellar disc. There is growing evidence that 
the stripping process, and the following quenching of the star formation activity, is more rapid ($\lesssim$ 0.5-1 Gyr; e.g.
Boselli et al. 2006, 2016a, 2021; Fossati et al. 2018) than the typical
crossing time of the cluster ($\simeq$ 1.7 Gyr in Virgo, Boselli \& Gavazzi 2006). The H$\alpha$ emission, which is sensitive to the emission of young 
($\lesssim$ 10 Myr) and massive O - early-B stars, stops $\simeq$ 10 Myr after any star formation activity is halted, and this occurs 
when all the ISM of the perturbed galaxies is removed during the interaction. In dwarf systems, this occurs on short timescales. Galaxies 
become quiescent systems and are no longer sampled in the H$\alpha$ luminosity function (flatter $\alpha$). In the NUV band, where the emission is dominated by 
A-F stars whose presence on the main sequence lasts $\simeq$ 100-500 Myr (Boissier et al. 2008; Boselli et al. 2009), galaxies are no longer counted in the luminosity function 
on longer timescales. They change colour, become red systems, and steepen the faint end of the luminosity function in the optical bands
at longer wavelengths. In massive systems, some gas can be retained in the inner regions where the gravitational potential well is the deepest.
Here, star formation can still occur, but at an overall rate that is slower than that of similar objects in the field.
A typical example is NGC 4569, a star-forming massive galaxy in the core of the Virgo cluster that goes through a ram-pressure stripping event and
is characterised by a truncated gaseous (HI and molecular) and star-forming disc (e.g. Vollmer et al. 2004; Boselli et al. 2006, 2016b).
Perturbed massive objects are indeed expected to reduce their activity, falling below the main-sequence relation (Boselli et al. 2023).
They thus shift to lower H$\alpha$ luminosities in the luminosity function, and might explain the observed decrease in the characteristic
H$\alpha$ luminosity observed in Virgo (log $L(H\alpha)^*$ = 41.49$^{41.60}_{41.39}$ erg s$^{-1}$) with respect to the field 
(41.74 $\leq$ log $L(H\alpha)^*$ $\leq$ 42.43 erg s$^{-1}$). The steep slope in the M and LVC clouds might  be caused by the first infall of these substructures
into the cluster. Galaxies that are predominantly HI-rich are just starting to feel the perturbation of the surrounding cluster environment (LVC)
here or are still too far away to feel it (M cloud, located at $\simeq$ 15 Mpc behind cluster A). They might just be galaxies that are pre-processed within their own 
substructure.
 
\begin{figure}
\centering
\includegraphics[width=0.5\textwidth]{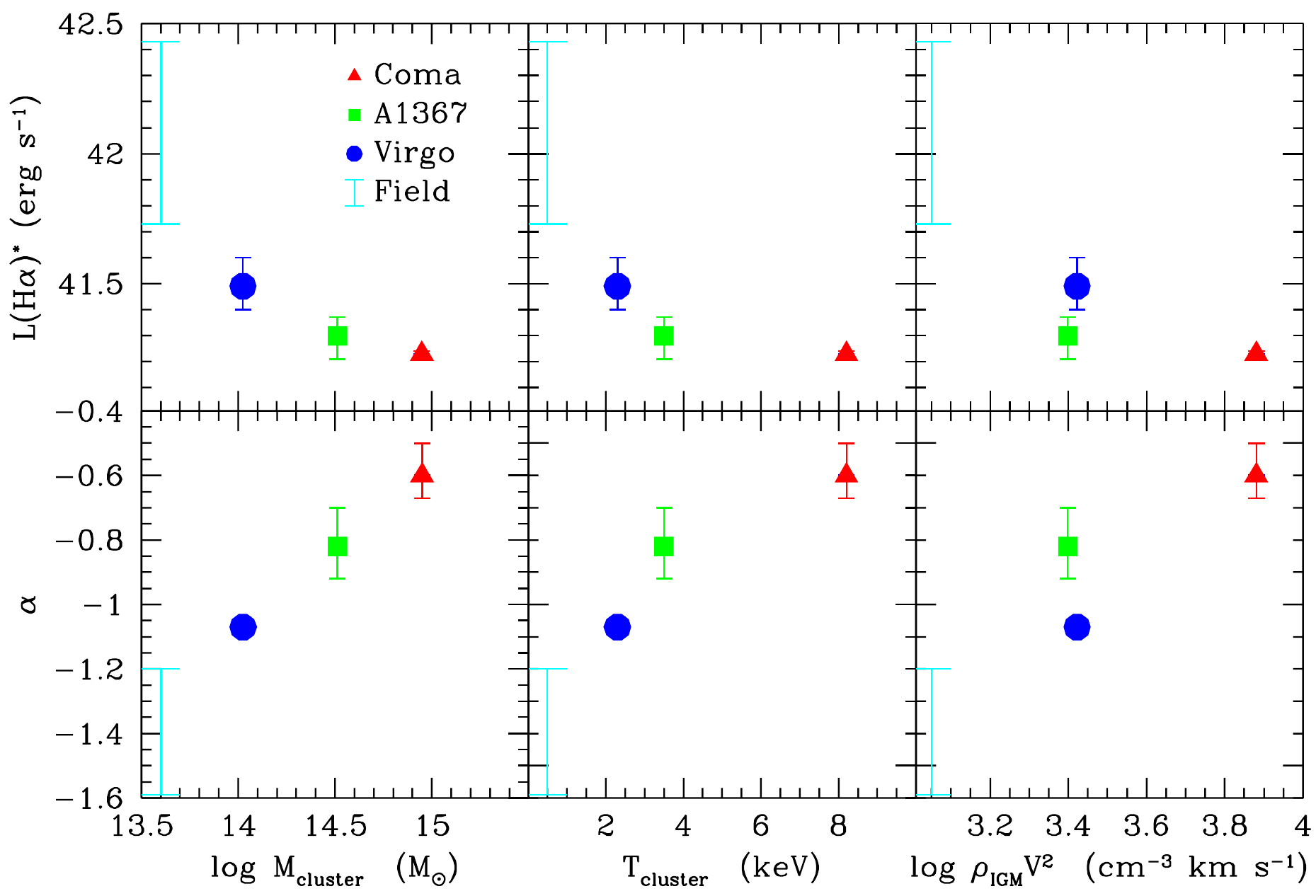}
\caption{Relation of the characteristic luminosity (upper row) and the slope (lower row) of the H$\alpha$ luminosity function 
derived in Virgo (this work), Coma, A1367 (Iglesias-Paramo et al. 2002), and the range of values derived for the field as a function
of the total dynamical mass of the clusters ($M_{200}$, from Boselli et al. 2022a, left column), the temperature of the intergalactic medium (central
column), and the mean pressure exerted by the intergalactic medium of density $\rho_{IGM}$ on galaxies moving within it at a velocity $V$ (right column).
The temperature of the gas, $\rho_{IGM}$, and $V$ are taken from Boselli \& Gavazzi (2006). The velocity is the mean velocity of the infalling late-type
galaxy population.
}
\label{RPS}%
\end{figure}

To determine how this evolutionar picture might depend on the cluster properties, we compare the best-fit parameters of the 
Schechter function derived for Coma, A1367, and Virgo in Fig. \ref{RPS}, where we plot the dependence
of the slope of the faint end $\alpha$ and the characteristic luminosity $L(H\alpha)^*$ (in log scale) as a function of three representative
cluster parameters, that is, the total mass of the clusters ($M_{200}$, taken from Boselli et al. 2022a), the typical temperature of the intergalactic medium, 
and the typical hydrodynamic pressure exerted by the intergalactic medium ($\rho_{IGM}$) on galaxies moving within the cluster with a mean velocity $V$.
For this purpose, we assumed the typical density of the intergalactic medium and the velocity dispersion of the three clusters (for the 
infalling spiral component) from Boselli \& Gavazzi (2006). Figure \ref{RPS} suggests that both the characteristic luminosity and 
the slope of the H$\alpha$ luminosity function are related to the properties of the clusters, and they increasingly deviate from those measured
in the field with the increase in the cluster mass, temperature, and exerted pressure ($P_{RPS}$ = $\rho_{IGM}V^2$). Although to be taken with caution
for the reasons mentioned above, Fig. \ref{RPS} is perfectly consistent with the evolutionary picture just mentioned, simply by considering 
that the gas-stripping and star formation-quenching phenomena become increasingly important in more extreme environments.  

\subsection{Comparison with high-$z$ clusters}

The comparison of the H$\alpha$ luminosity function derived in this work with the one determined for clusters at intermediate and high redshift 
is of paramount importance for identifying the main perturbing mechanisms in galaxy evolution with cosmic time.
In particular, comparing the statistical properties of samples of cluster galaxies selected according to similar criteria at different epochs 
is crucial for posing strong observational constraints on the timescales for the quenching phenomenon that is now well established in nearby cluster galaxies.
Considerable effort has been expended in recent years to derive the star formation properties of galaxies in clusters at different redshift 
using the H$\alpha$ emission line (e.g. Finn et al. 2008; Nantais et al. 2020). In several cases, H$\alpha$ selected galaxies were used to do this, either using the grism spectroscopy provided by the HST (e.g. Zeimann et al. 2013; Cooper et al. 2022), 
or using narrow-band imaging data such as those gathered in this work (e.g. Kodama et al. 2004; Finn et al. 2004, 2005; 
Koyama et al. 2010, 2011, 2018; Sobral et al. 2011, 2016; Shimakawa et al 2018). 
Overall, these works consistently indicate that i) the star formation activity in the densest regions of rich clusters is suppressed
with respect to the field, and ii) that the mean star formation activity of cluster galaxies rapidly decreases from $z$ $\simeq$ 0.8 to $z$ $\simeq$ 0.
The H$\alpha$ luminosity function of intermediate-redshift cluster galaxies (0.4 $\lesssim$ $z$ $\lesssim$ 0.8), however, was derived only for a very
limited number of objects. The results obtained from these analyses are not always consistent, possibly because of poor statistics, a 
strong cluster mass dependence, and rapid galaxy evolution. 
A comparison of the best-fit Schechter parameters derived in these works with those of Virgo is made difficult by the very different
sampled H$\alpha$ luminosity range, which is limited to $L(H\alpha)$ $\gtrsim$ 10$^{40}$ erg s$^{-1}$ for clusters at $z$ $\simeq$ 0.4 and 
$L(H\alpha)$ $\gtrsim$ 10$^{41}$ erg s$^{-1}$ for clusters at $z$ $\simeq$ 0.8. According to the well-known covariance of the Schechter function parameters 
($\alpha$ and $L^*$ are correlated variables which depend on the sampled luminosity range), the observed flattening of the H$\alpha$ luminosity function observed in Virgo 
in the range 10$^{38}$ $\lesssim$ $L(H\alpha)$ $\lesssim$ 10$^{41}$ erg s$^{-1}$ cannot be measured at higher redshift if, as expected, $L^*(H\alpha)$
is close to the detection limit.
Keeping in mind all these possible uncertainties, we can speculate that the
decrease in the star formation activity of cluster galaxies occurred in the last $<$ 4 Gyr principally in dwarf systems.
This statement is suggested by the recent build-up of the faint end of the red sequence observed in clusters 
(e.g. De Lucia et al. 2007, 2009; Stott et al. 2007, 2009; Gilbank \& Balogh 2008)
and favours an efficient mechanism that can remove the gas from the galactic disc and quench the activity of star formation on short timescales
(e.g. ram-pressure stripping; Boselli et al. 2022a).

\begin{table}
\caption{Properties of the Virgo-like clusters selected in the TNG simulations at $z$=0, in comparison to the VESTIGE observations}
\label{TNG}
{
\[
\begin{tabular}{ccccc}
\hline
\noalign{\smallskip}
\hline
TNG     & FoFID         & Halo mass                     & N. member galaxies    & SFR limit                       \\
        &               & $\times$ 10$^{14}$ M$_{\odot}$&               & M$_{\odot}$ yr$^{-1}$   \\
\hline  
50      & 0             & 1.8                           &50             & 10$^{-5}$       \\
50      & 1             & 0.9                           &59             & 10$^{-5}$       \\
100     & 0             & 3.6                           &101            & 10$^{-4}$       \\
100     & 1             & 3.7                           &38             & 10$^{-4}$       \\
100     & 2             & 3.3                           &25             & 10$^{-4}$       \\
100     & 3             & 1.7                           &61             & 10$^{-4}$       \\
100     & 4             & 2.5                           &43             & 10$^{-4}$       \\
100     & 5             & 2.0                           &36             & 10$^{-4}$       \\
100     & 6             & 2.0                           &11             & 10$^{-4}$       \\
100     & 8             & 2.0                           &37             & 10$^{-4}$       \\
100     & 9             & 2.1                           &77             & 10$^{-4}$       \\
100     & 10            & 1.6                           &21             & 10$^{-4}$       \\
100     & 11            & 1.3                           &13             & 10$^{-4}$       \\
100     & 14            & 1.1                           &13             & 10$^{-4}$       \\
100     & 15            & 1.1                           &18             & 10$^{-4}$       \\
100     & 17            & 1.0                           &23             & 10$^{-4}$       \\
Virgo   & -             & 1.0-4.0                       &384            & 10$^{-5}$       \\
\noalign{\smallskip}
\hline
\end{tabular}
\]
}
\end{table}

\subsection{Comparison with cosmological simulations}

\begin{figure}
\centering
\includegraphics[width=0.5\textwidth]{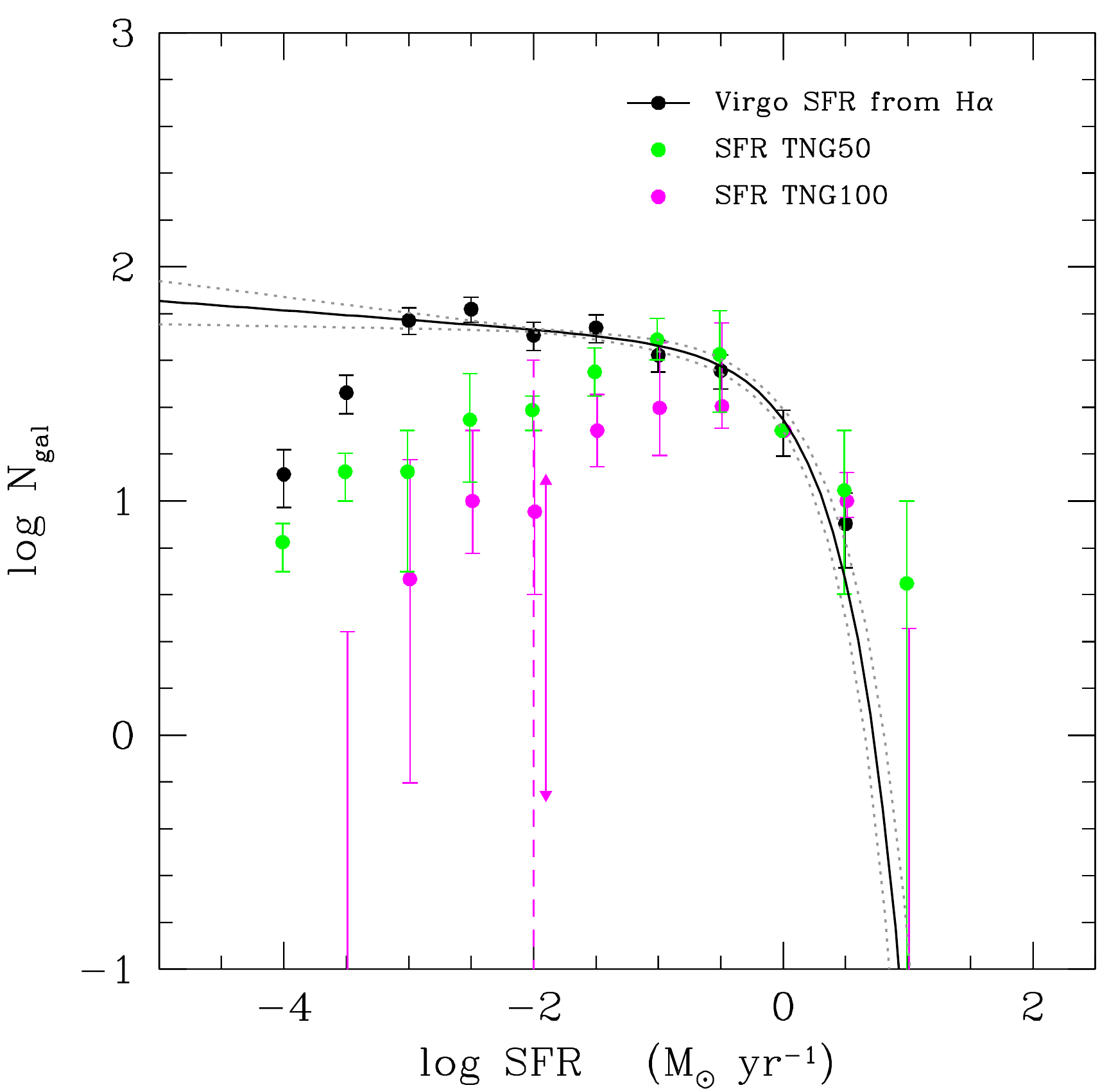}
\caption{Comparison of the $SFR$ luminosity function of the Virgo cluster derived using the VESTIGE H$\alpha$ 
imaging data (solid and dotted lines, representing the 1$\sigma$ confidence interval) and the luminosity functions
derived from the TNG50 (green symbols) and TNG100 (magenta symbols) simulated clusters and the galaxies therein.
Both TNG luminosity functions are normalised to the number of galaxies within the Virgo 
cluster sample in the bin centred at $SFR$ = 1 M$_{\odot}$ yr$^{-1}$. The error bar in the TNG50 data denotes the 
dynamic range covered when considering the two simulated massive clusters as separate entities. The error bar in the TNG100
curve shows the 25\%\ and 75\%\ percentiles across the 14 simulated hosts. The vertical dashed magenta line at $SFR$ = 10$^{-2}$ 
M$_{\odot}$ yr$^{-1}$ gives the dynamic range covered by the 14 simulated clusters, and the vertical double arrow denotes 
the systematic uncertainties in the way the SFR can be estimated from the simulated cluster members.
}
\label{LFTNG}%
\end{figure}

In Fig. \ref{LFTNG} we compare the SFR function of the Virgo cluster presented in Fig. \ref{LFHafieldcluster} with the 
predictions of hydrodynamical cosmological simulations of galaxies, specifically, those from the IllustrisTNG project\footnote{\url{https://www.tng-project.org}} 
(Nelson et al. 2019). These are based on an extensive model of galaxy formation, including the solution of gravity and magnetohydrodynamics 
in expanding universes and feedback from stars and from supermassive black holes (Pillepich et al. 2018a), so that phenomena that drive 
environmental processes in dense groups and clusters (tidal and ram-pressure stripping, gravitational interactions, mergers, etc.) 
all emerge from the solution of the numerically solved equations.

For this purpose, we used two different sets of simulated clusters and the galaxies therein. The first set was extracted from 
the smallest volume and highest resolution TNG50 run (Pillepich et al. 2019; Nelson et al. 2019), which allowed us to sample
star-forming galaxies down to the stellar mass limit of $M_{star}$ $\simeq$ 10$^6$ M$_{\odot}$ and $SFR$ $\simeq$ 10$^{-5}$ 
M$_{\odot}$ yr$^{-1}$ (Joshi et al. 2021), comparable to the stellar mass and star-forming range sampled by the VESTIGE survey (Boselli et al. 2023). 
The TNG50 simulation includes two clusters of mass $M_{cluster}$ $\simeq$ 10$^{14}$ M$_{\odot}$ (extensively studied e.g. by Joshi et al. (2020)), comparable to the 
mass of the Virgo cluster. These two massive dark matter haloes have a total of 109 star-forming galaxies within $r_{200}$
(projected distance, see Table \ref{TNG}). To increase the statistics and assess the effects of variance, we also considered the TNG100 run
(Naiman et al. 2018; Marinacci et al. 2018; Nelson et al. 2018; Pillepich et al. 2018b; Springel et al. 2018),
which includes 14 clusters of mass 10$^{14}$ $\leq$ $M_{cluster}$ $\leq$ 4 $\times$ 10$^{14}$ M$_{\odot}$. For these
simulations, the limiting SFR is $SFR$ $\simeq$ 10$^{-4}$ M$_{\odot}$ yr$^{-1}$ , while the one
in stellar mass is rather $M_{star}$ $\simeq$ 10$^8$ M$_{\odot}$ (Donnari et al. 2019). The TNG100 clusters 
used in this analysis include 517 star-forming galaxies, that is, galaxies with a non-vanishing simulated SFR.

We identified galaxy members of the simulated clusters using the seme criteria as were applied in the case of the Virgo cluster, that is, as those 
objects located at a projected distance $R$ $\leq$ $r_{200}$ (which corresponds to the angular extension of the VESTIGE 
footprint) and with a line-of-sight (LoS) velocity with respect to the cluster $v_{LoS}$ $\leq$ 2000 km s$^{-1}$,
the velocity range covered by the NB filter (see Boselli et al. 2018a 
for details). For the simulated galaxies, this corresponds to the subhalo velocity relative to the host in the z-direction,
here taken as a proxy for the LoS velocity. 

To focus only on the shapes of the different luminosity functions, the results of the simulations were normalised to the number of galaxies observed in Virgo
within the star-forming rate bin centred at $SFR$ = 1 M$_{\odot}$ yr$^{-1}$. The comparison at high SFRs
($SFR$ $\gtrsim$ 10 M$_{\odot}$ yr$^{-1}$) should be considered with extreme caution given the very limited number of observed 
and simulated objects. We thus focus on the slope of the distribution for $SFR$ $\leq$ 1 M$_{\odot}$ yr$^{-1}$.

While the observed Virgo shows a mild but significant increase in the number of galaxies with a decreasing SFR in the range
10$^{-3}$ $\lesssim$ $SFR$ $\lesssim$ 1 M$_{\odot}$ yr$^{-1}$ ($\alpha$ = -1.04), the number of star-forming objects 
decreases with decreasing SFR in the TNG50 and TNG100 simulated clusters, that is, the drop in number of 
star-forming galaxies seems to occur at lower SFR values in the simulations than in the Virgo cluster. This decrease is more 
pronounced in the TNG100 simulated clusters than in TNG50 clusters: Even though the quenched fractions of Virgo-mass cluster galaxies 
have been shown to be completely robust against resolution effects between TNG100 and TNG50 (Donnari et al. 2021b, see their Fig. 10), 
some resolution effects might affect the simulation-predicted actual values of SFR, especially at these low levels of star formation. 
For simulated galaxies, there can also be large systematic uncertainties or differences in the way in which the galaxy 
SFRs are estimated: The thick vertical magenta error bar at $SFR$ = 10$^{-2}$ M$_{\odot}$ yr$^{-1}$ shows an estimate 
of this uncertainty, which amounts to 0.6 dex at least.

Barring resolution effects and systematic issues in the sample selection and in the comparison of the SFR proxies, 
the difference between the $SFR$ luminosity functions of the TNG50 and TNG100 simulations on the one hand and that observed in Virgo, 
on the other, might have several origins. The error bar in the TNG50 curve of Fig. \ref{LFTNG} indicates the dynamic range of the luminosity function when the two massive clusters are considered as 
separate entities. The error bar in the TNG100 luminosity function gives the 25\%\ and 75\%\ percentiles of the distributions of the 14 clusters when 
they are considered as independent entities. According to the TNG simulations and based on the luminosity functions across 16 hosts, the cluster-to-cluster 
variation may be  0.9 dex at least at $SFR$ = 10$^{-2}$ M$_{\odot}$ yr$^{-1}$. This is larger than the difference between the observed $SFR$ 
luminosity function of Virgo and the average simulated luminosity functions (about $0.4-0.7$ dex at the same $SFR$ value). We note that despite the large 
dispersion in the number of star-forming objects per luminosity bin measured in the 14 individual TNG100 clusters, none of them reaches 
the values observed in Virgo for $SFR<10^{-2}$ , and all show a decreasing number of objects going from $SFR<10^{-1}$ to $SFR<10^{-3}$ M$_{\odot}$ yr$^{-1}$.

At the same time, Virgo might not be representative of clusters with a dynamical mass 10$^{14}$ $\lesssim$ $M_{cluster}$ $\lesssim$ 4 $\times$ 10$^{14}$ M$_{\odot}$ ($M_{200}$) 
because it is a young dynamically unrelaxed system that is still populated by gas-rich star-forming objects when compared to other nearby clusters 
(e.g. Boselli \& Gavazzi 2006, Boselli et al. 2014). To examine this possibility, the observation of 
other nearby clusters of similar mass is paramount, and in the context of validating the realism of simulations, a comparison with a larger number 
of simulated Virgo-like hosts is of the essence. 

Despite the complexities discussed above, the comparison of Fig. \ref{LFTNG} may suggest that the lower number and the decrease in the number of star -forming objects at the faint end of the luminosity function in TNG Virgo-like simulated clusters may be related to the efficiency of the various 
physical mechanisms in removing the gas from member galaxies (ram-pressure stripping, gravitational perturbations, and feedback). Similar comparisons 
have been made extensively in previous analyses: For example, Donnari et al. (2021b) have shown that TNG produces quenched fractions 
for both centrals and satellites that are broadly consistent with observations at $z\sim0$, at least for galaxies more massive than 
$M_{star}$ $\simeq$ 10$^9$ M$_{\odot}$, even though the quantitative agreement between TNG and the Sloan Digital Sky Survey 
for satellite quenched fractions in groups and clusters strongly depends on the host and galaxy mass range. 
For cluster galaxies, the level of agreement was found to be strikingly good for intermediate group-mass scale hosts (10$^{13-14}$ M$_{\odot}$) 
and to decrease for TNG low-mass galaxies in either Milky Way-mass or massive (10$^{15}$ M$_{\odot}$) hosts. 

\begin{figure*}
\centering
\includegraphics[width=0.49\textwidth]{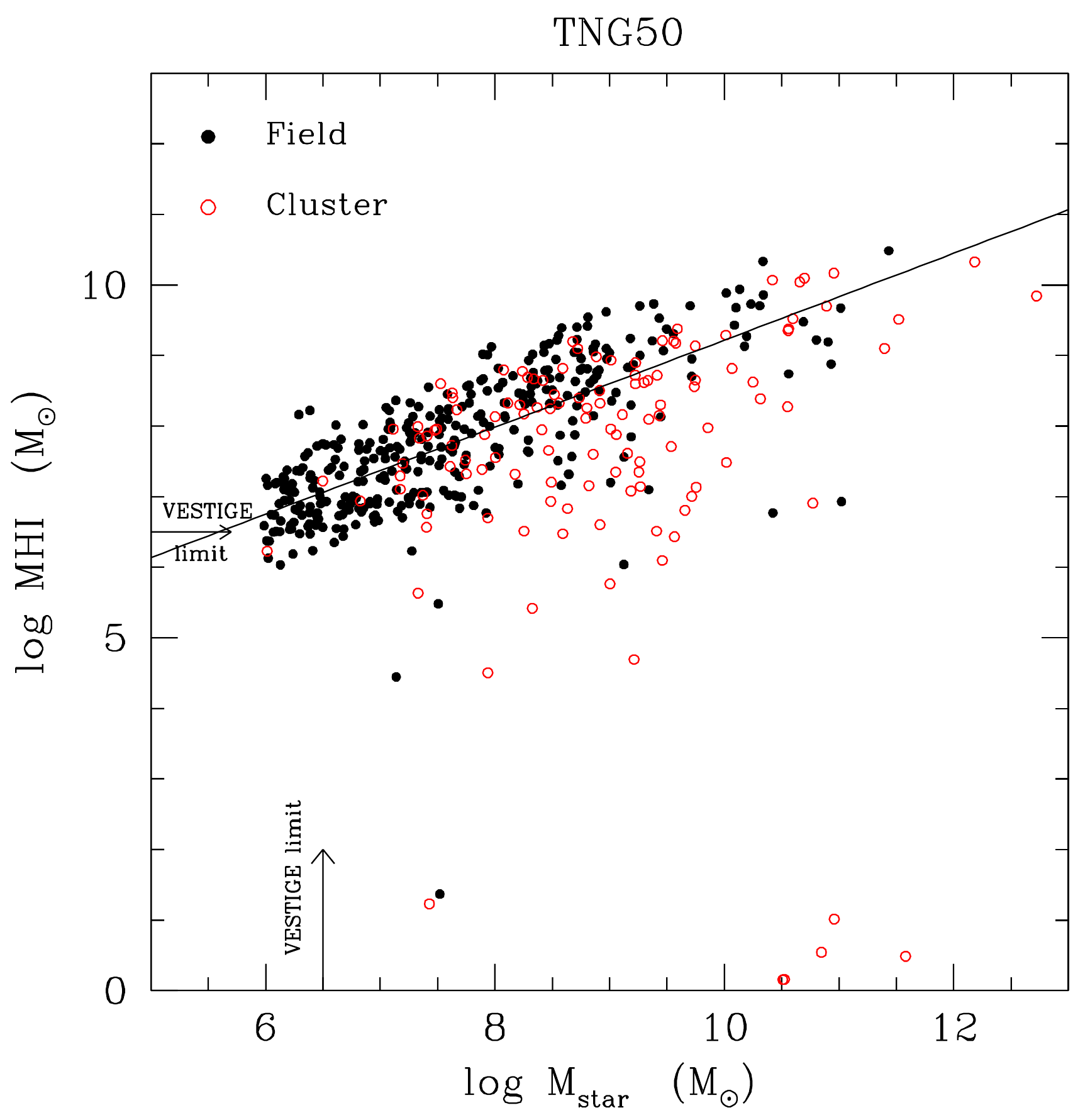}
\includegraphics[width=0.49\textwidth]{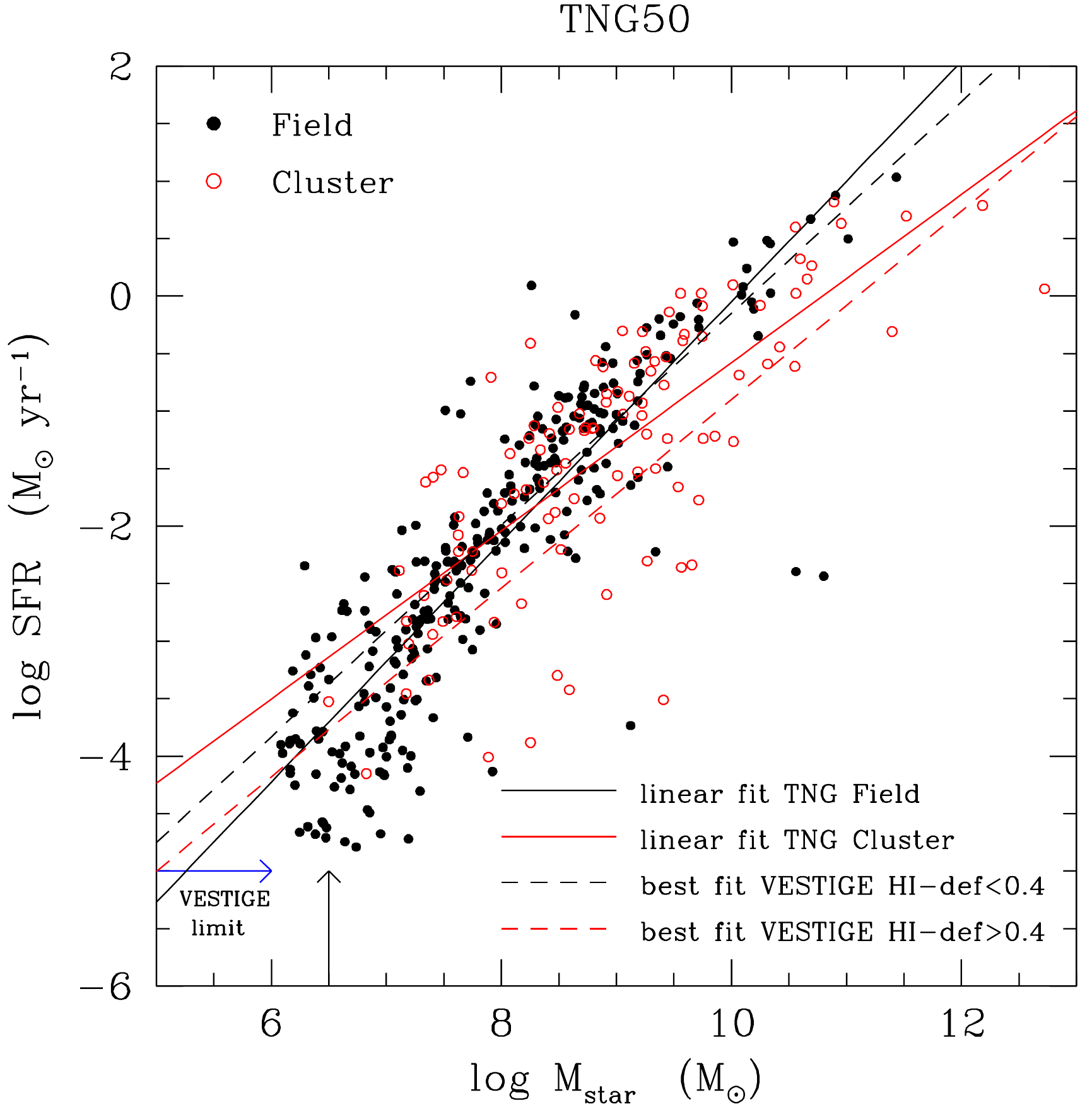}\\
\caption{Scaling relation of the atomic gas mass and the stellar mass (left panel) and the SFR and 
the stellar mass (right panel) for galaxies extracted from the TNG50 simulations. Red empty circles show galaxies within
the two massive clusters, and filled black dots show galaxies in the field ($R$ $>$ 2 $\times$ $r_{200}$). The solid lines
in the two panels show the linear fit for field (black) and cluster (red) galaxies, and the dashed lines in the right panel show the best fit for gas-rich
(black) and gas-poor (red) galaxies in the Virgo cluster derived by Boselli et al. (2023). The black arrows indicate 
the approximate limit in stellar mass and HI gas mass for the H$\alpha$-selected Virgo sample, and the blue arrow shows the
limiting sensitivity in $SFR$ for the VESTIGE data. 
}
\label{scaling}%
\end{figure*}

As expected from theoretical arguments and observations, galaxies in dense environments exhibit a lower gas-mass content than their analogues in 
the field: We refer to the simulated cluster objects in the $M_{HI}$ versus $M_{star}$ and  $SFR$ versus $M_{star}$ scaling relations in Fig. \ref{scaling}, for instance. 
In a previous detailed analysis, Stevens et al. (2019) have shown that the HI content of TNG100 galaxies is aligned with that inferred from 
surveys such as ALFALFA (Giovanelli et al. 2005) and the extended GALEX Arecibo SDSS survey (xGASS, Catinella et al. 2018). This is also the case here, 
where the simulated galaxies follow the scaling relations traced by the Virgo objects.
There is, however, a sistematic difference in the low-mass regime that is not sampled in the xGASS sample, which is limited to $M_{star}$ $>$ 10$^9$ M$_{\odot}$.
Figure \ref{scaling} shows that in the stellar mass range $M_{star}$ $\lesssim$ 10$^{8}$ M$_{\odot}$ , the number of simulated cluster objects 
in the $M_{HI}$ versus $M_{star}$ and $SFR$ versus $M_{star}$ scaling relations is significantly smaller than that in the field. 
Most of the simulated dwarf galaxies in the cluster do not contain cold gas, 
thus their star formation activity is completely stopped, suggesting that the stripping process in these systems is too efficient.
Although the observed catalogue is H$\alpha$ selected, we can compare the fraction 
of star-forming objects at a stellar mass limit in the observations and in the simulations. The NGVS catalogue is complete down to 
a stellar mass of 1.6 $\times$ 10$^7$ M$_{\odot}$ (1340 objects catalogued as Virgo cluster members with a stellar mass above this limit; Ferrarese et al. 2020).
At this stellar mass limit, the VESTIGE survey detected 360 star forming galaxies (27\%), while the fraction of star-forming 
objects in the TNG50 simulated clusters is 103/987 = 10\%. These considerations suggest that if Virgo is an average host, the stripping process in the TNG simulations may be too efficient for very low-mass satellites in Virgo-mass clusters.

\section{Conclusion}

Based on their untargeted nature, we used the NB H$\alpha$ imaging data for 384 galaxies gathered during the VESTIGE survey to derive 
the first H$\alpha$ luminosity function of the galaxies in the Virgo cluster within its virial radius. The extremely deep data allowed us to measure 
the H$\alpha$ emission of galaxies from the brightest objects down to the emission of a single ionising star, thus sampling the full 
range of the H$\alpha$ luminosity function (10$^{36}$ $\leq$ $L(H\alpha)$ $\leq$ 10$^{42}$ erg s$^{-1}$) throughout the cluster.
The H$\alpha$ luminosity function was derived for the whole cluster, for its different substructures, and separately for HI-rich and HI-poor
galaxies, with the latter considered as systems perturbed by their surrounding environment.
Using a unique set of multifrequency data, we corrected the observed H$\alpha$ luminosities for [NII] contamination and dust attenuation, and 
converted them to derive the SFR function of the cluster members. We fit the H$\alpha$ and $SFR$ luminosity functions
with a Schechter function. These luminosity functions steeply increase up to their characteristic luminosity 
($L(H\alpha)^*$ = 10$^{41.49}$ erg s$^{-1}$, $SFR^*$ = 10$^{0.15}$ M$_{\odot}$ yr$^{-1}$), have a flat slope ($\alpha_{H\alpha}$ = -1.07 
and $\alpha_{SFR}$ = -1.03) in the range 10$^{38.5}$ $\lesssim$ $L(H\alpha)$ $\lesssim$ 10$^{40.5}$ erg s$^{-1}$ and 
10$^{-3}$ $\lesssim$ $SFR$ $\lesssim$ 10$^{-0.5}$ M$_{\odot}$ yr$^{-1}$, and abruptly decline at lower luminosities.

The IllustrisTNG cosmological hydrodynamical simulations for clusters with a mass similar to that of Virgo (1-4 $\times$ 10$^{14}$ M$_{\odot}$) 
predict a number of galaxies that is lower than is observed by VESTIGE for Virgo by 0.2-0.6 dex with TNG50 and 0.5-1.0 
dex with TNG100 in the SFR range 10$^{-3}$ $\leq$ $SFR$ $\leq$ 0.1 M$_{\odot}$ yr$^{-1}$. 
Although a large ($\simeq$ 0.9 dex) cluster-to-cluster variation is predicted by the simulations and the significant uncertainty 
related to the use of different tracers of star formation ($\simeq$ 0.6 dex), none of the 16 simulated clusters reproduces the 
SFR function observed in Virgo.
If Virgo is representative of clusters of this mass, this difference suggests that 
gas stripping is more efficient in the simulated low-mass satellites in Virgo-mass clusters than observed.
Only the observations of other nearby clusters with similar properties can solve this important question.

When compared to those derived for the field and for Coma and A1367, the characteristic parameters of the fitted Schechter function 
of the observed H$\alpha$ luminosity function seem related with the total dynamical mass of the cluster, the temperature of the 
of the hot IGM, and the mean dynamical pressure exerted by the IGM on the galaxies moving within the cluster. All these relations,
which need to be confirmed on a larger statistical sample of local clusters of different mass, suggest that
the activity of star formation of galaxies is reduced with increasing efficiency in more massive clusters. This result is expected 
by all modern cosmological numerical simulations, including IllustrisTNG (Donnari et al. 2021a), that is, in a scenario
in which the dominant mechanism that causes the gas stripping and the following quenching of the star formation activity in massive clusters
($M_{cluster}$ $\gtrsim$ 10$^{14}$ M$_{\odot}$) is the hydrodynamic interaction of the ISM of the infalling star-forming galaxies with 
the hot IGM trapped with the gravitational potential well of the cluster (ram-pressure stripping).  

\begin{acknowledgements}

We thank the anonymous referee for constructive comments and suggestions which helped improving the quality of the manuscript.
We are grateful to the whole CFHT team who assisted us in the preparation and in the execution of the observations and in the calibration and data reduction: 
Todd Burdullis, Daniel Devost, Bill Mahoney, Nadine Manset, Andreea Petric, Simon Prunet, Kanoa Withington.
We acknowledge financial support from ``Programme National de Cosmologie and Galaxies" (PNCG) funded by CNRS/INSU-IN2P3-INP, CEA and CNES, France.
This research has made use of the NASA/IPAC Extragalactic Database (NED) 
which is operated by the Jet Propulsion Laboratory, California Institute of 
Technology, under contract with the National Aeronautics and Space Administration
and of the GOLDMine database (http://goldmine.mib.infn.it/) (Gavazzi et al. 2003).
MA and DN acknowledge funding from the Deutsche Forschungsgemeinschaft (DFG) through an 
Emmy Noether Research Group (grant number NE 2441/1-1).
MB gratefully acknowledges support by the ANID BASAL project FB210003 and from the FONDECYT regular grant 1211000.

\end{acknowledgements}


\begin{thebibliography}{}

\bibitem[Baldwin et al.(1981)]{1981PASP...93....5B} Baldwin, J.~A., Phillips, M.~M., \& Terlevich, R.\ 1981, \pasp, 93, 5 
\bibitem[Balogh \& Morris(2000)]{2000MNRAS.318..703B} Balogh, M.~L. \& Morris, S.~L.\ 2000, \mnras, 318, 703
\bibitem[Balogh et al.(2002)]{2002MNRAS.335...10B} Balogh, M.~L., Couch, W.~J., Smail, I., et al.\ 2002, \mnras, 335, 10
\bibitem[Bellhouse et al.(2019)]{2019MNRAS.485.1157B} Bellhouse, C., Jaff{\'e}, Y.~L., McGee, S.~L., et al.\ 2019, \mnras, 485, 1157
\bibitem[Benson et al.(2003)]{2003ApJ...599...38B} Benson, A.~J., Bower, R.~G., Frenk, C.~S., et al.\ 2003, \apj, 599, 38
\bibitem[Bertin \& Arnouts(1996)]{1996A&AS..117..393B} Bertin, E., \& Arnouts, S.\ 1996, \aaps, 117, 393 
\bibitem[Bialas et al.(2015)]{2015A&A...576A.103B} Bialas, D., Lisker, T., Olczak, C., et al.\ 2015, \aap, 576, A103
\bibitem[Binggeli et al.(1985)]{1985AJ.....90.1681B} Binggeli, B., Sandage, A., \& Tammann, G.~A.\ 1985, \aj, 90, 1681 
\bibitem[Binggeli et al.(1987)]{1987AJ.....94..251B} Binggeli, B., Tammann, G.~A., \& Sandage, A.\ 1987, \aj, 94, 251 
\bibitem[Blanton et al.(2003)]{2003ApJ...592..819B} Blanton, M.~R., Hogg, D.~W., Bahcall, N.~A., et al.\ 2003, \apj, 592, 819
\bibitem[Blanton et al.(2005)]{2005ApJ...631..208B} Blanton, M.~R., Lupton, R.~H., Schlegel, D.~J., et al.\ 2005, \apj, 631, 208
\bibitem[B{\"o}hringer et al.(1994)]{1994Natur.368..828B} B{\"o}hringer, H., Briel, U.~G., Schwarz, R.~A., et al.\ 1994, \nat, 368, 828 
\bibitem[Boissier et al.(2008)]{2008ApJ...681..244B} Boissier, S., Gil de Paz, A., Boselli, A., et al.\ 2008, \apj, 681, 244-257 
\bibitem{2006PASP..118..517B} Boselli, A., \& Gavazzi, G.\ 2006, \pasp, 118, 517 
\bibitem[Boselli \& Gavazzi(2014)]{2014A&ARv..22...74B} Boselli, A., \& Gavazzi, G.\ 2014, \aapr, 22, 74 
\bibitem{2008ApJ...674..742B} Boselli, A., Boissier, S., Cortese, L., \& Gavazzi, G.\ 2008, \apj, 674, 742 
\bibitem[Boselli et al.(2009)]{2009ApJ...706.1527B} Boselli, A., Boissier, S., Cortese, L., et al.\ 2009, \apj, 706, 1527 
\bibitem{2011A&A...528A.107B} Boselli, A., Boissier, S., Heinis, S., et al.\ 2011, \aap, 528, A107 
\bibitem[Boselli et al.(2014)]{2014A&A...570A..69B} Boselli, A., Voyer, E., Boissier, S., et al.\ 2014, \aap, 570, AA69 
\bibitem[Boselli et al.(2015)]{2015A&A...579A.102B} Boselli, A., Fossati, M., Gavazzi, G., et al.\ 2015, \aap, 579, A102 
\bibitem[Boselli et al.(2016)]{2016A&A...596A..11B} Boselli, A., Roehlly, Y., Fossati, M., et al.\ 2016a, \aap, 596, A11 
\bibitem[Boselli et al.(2016)]{2016A&A...587A..68B} Boselli, A., Cuillandre, J.~C., Fossati, M., et al.\ 2016b, \aap, 587, A68 
\bibitem[Boselli et al.(2016)]{2016A&A...585A...2B} Boselli, A., Boissier, S., Voyer, E., et al.\ 2016c, \aap, 585, A2 
\bibitem[Boselli et al.(2018)]{2018A&A...614A..56B} Boselli, A., Fossati, M., Ferrarese, L., et al.\ 2018a, \aap, 614, A56 
\bibitem[Boselli et al.(2018)]{2018A&A...620A.164B} Boselli, A., Fossati, M., Consolandi, G., et al.\ 2018b, \aap, 620, A164      
\bibitem[Boselli et al.(2019)]{2019A&A...623A..52B} Boselli, A., Fossati, M., Longobardi, A., et al.\ 2019, \aap, 623, A52 
\bibitem[Boselli et al.(2021)]{2021A&A...646A.139B} Boselli, A., Lupi, A., Epinat, B., et al.\ 2021, \aap, 646, A139
\bibitem[Boselli et al.(2022)]{2022A&ARv..30....3B} Boselli, A., Fossati, M., \& Sun, M.\ 2022a, \aapr, 30, 3
\bibitem[Boselli et al.(2022)]{2022A&A...659A..46B} Boselli, A., Fossati, M., Longobardi, A., et al.\ 2022b, \aap, 659, A46
\bibitem[Boselli et al.(2023)]{2023A&A...669A..73B} Boselli, A., Fossati, M., Roediger, J., et al.\ 2023, \aap, 669, A73
\bibitem[Bothwell et al.(2011)]{2011MNRAS.415.1815B} Bothwell, M.~S., Kennicutt, R.~C., Johnson, B.~D., et al.\ 2011, \mnras, 415, 1815
\bibitem[Byrd \& Valtonen(1990)]{1990ApJ...350...89B} Byrd, G., \& Valtonen, M.\ 1990, \apj, 350, 89 
\bibitem[Calzetti et al.(2010)]{2010ApJ...714.1256C} Calzetti, D., Wu, S.-Y., Hong, S., et al.\ 2010, \apj, 714, 1256 
\bibitem[Cardelli et al.(1989)]{1989ApJ...345..245C} Cardelli, J.~A., Clayton, G.~C., \& Mathis, J.~S.\ 1989, \apj, 345, 245
\bibitem[Catinella et al.(2018)]{2018MNRAS.476..875C} Catinella, B., Saintonge, A., Janowiecki, S., et al.\ 2018, \mnras, 476, 875
\bibitem[Cattorini et al.(2023)]{2023A&A...671A.118C} Cattorini, F., Gavazzi, G., Boselli, A., et al.\ 2023, \aap, 671, A118
\bibitem[Chisholm et al.(2018)]{2018A&A...616A..30C} Chisholm, J., Gazagnes, S., Schaerer, D., et al.\ 2018, \aap, 616, A30
\bibitem[Choi et al.(2020)]{2020ApJ...902...54C} Choi, Y., Dalcanton, J.~J., Williams, B.~F., et al.\ 2020, \apj, 902, 54
\bibitem[Cid Fernandes et al.(2011)]{2011MNRAS.413.1687C} Cid Fernandes, R., Stasi{\'n}ska, G., Mateus, A., et al.\ 2011, \mnras, 413, 1687
\bibitem[Cooper et al.(2022)]{2022MNRAS.509.5382C} Cooper, J.~R., Rudnick, G.~H., Brammer, G.~G., et al.\ 2022, \mnras, 509, 5382
\bibitem[Couch et al.(2001)]{2001ApJ...549..820C} Couch, W.~J., Balogh, M.~L., Bower, R.~G., et al.\ 2001, \apj, 549, 820
\bibitem[Cowie \& Songaila(1977)]{1977Natur.266..501C} Cowie, L.~L., \& Songaila, A.\ 1977, \nat, 266, 501 
\bibitem[Croton et al.(2006)]{2006MNRAS.365...11C} Croton, D.~J., Springel, V., White, S.~D.~M., et al.\ 2006, \mnras, 365, 11
\bibitem[Dale et al.(2010)]{2010ApJ...712L.189D} Dale, D.~A., Barlow, R.~J., Cohen, S.~A., et al.\ 2010, \apjl, 712, L189
\bibitem[De Lucia et al.(2007)]{2007MNRAS.374..809D} De Lucia, G., Poggianti, B.~M., Arag{\'o}n-Salamanca, A., et al.\ 2007, \mnras, 374, 809
\bibitem[De Lucia et al.(2009)]{2009MNRAS.400...68D} De Lucia, G., Poggianti, B.~M., Halliday, C., et al.\ 2009, \mnras, 400, 68
\bibitem[De Lucia et al.(2012)]{2012MNRAS.423.1277D} De Lucia, G., Weinmann, S., Poggianti, B.~M., Arag{\'o}n-Salamanca, A., \& Zaritsky, D.\ 2012, \mnras, 423, 1277 
\bibitem[De Propris et al.(2003)]{2003MNRAS.342..725D} De Propris, R., Colless, M., Driver, S.~P., et al.\ 2003, \mnras, 342, 725
\bibitem[de Vaucouleurs(1961)]{1961ApJS....6..213D} de Vaucouleurs, G.\ 1961, \apjs, 6, 213 
\bibitem[Donnari et al.(2019)]{2019MNRAS.485.4817D} Donnari, M., Pillepich, A., Nelson, D., et al.\ 2019, \mnras, 485, 4817
\bibitem[Donnari et al.(2021)]{2021MNRAS.500.4004D} Donnari, M., Pillepich, A., Joshi, G.~D., et al.\ 2021a, \mnras, 500, 4004
\bibitem[Donnari et al.(2021)]{2021MNRAS.506.4760D} Donnari, M., Pillepich, A., Nelson, D., et al.\ 2021b, \mnras, 506, 4760
\bibitem[Dressler(1980)]{1980ApJ...236..351D} Dressler, A.\ 1980, \apj, 236, 351 
\bibitem[Dressler et al.(1997)]{1997ApJ...490..577D} Dressler, A., Oemler, A., Jr., Couch, W.~J., et al.\ 1997, \apj, 490, 577 
\bibitem[Feroz \& Hobson(2008)]{2008MNRAS.384..449F} Feroz, F. \& Hobson, M.~P.\ 2008, \mnras, 384, 449
\bibitem[Feroz et al.(2019)]{2019OJAp....2E..10F} Feroz, F., Hobson, M.~P., Cameron, E., et al.\ 2019, The Open Journal of Astrophysics, 2, 10
\bibitem[Ferrarese et al.(2006)]{2006ApJS..164..334F} Ferrarese, L., C{\^o}t{\'e}, P., Jord{\'a}n, A., et al.\ 2006, \apjs, 164, 334 
\bibitem[Ferrarese et al.(2012)]{2012ApJS..200....4F} Ferrarese, L., C{\^o}t{\'e}, P., Cuillandre, J.-C., et al.\ 2012, \apjs, 200, 4 
\bibitem[Ferrarese et al.(2016)]{2016ApJ...824...10F} Ferrarese, L., C{\^o}t{\'e}, P., S{\'a}nchez-Janssen, R., et al.\ 2016, \apj, 824, 10 
\bibitem[Ferrarese et al.(2020)]{2020ApJ...890..128F} Ferrarese, L., C{\^o}t{\'e}, P., MacArthur, L.~A., et al.\ 2020, \apj, 890, 128
\bibitem[Finn et al.(2004)]{2004ApJ...604..141F} Finn, R.~A., Zaritsky, D., \& McCarthy, D.~W.\ 2004, \apj, 604, 141
\bibitem[Finn et al.(2005)]{2005ApJ...630..206F} Finn, R.~A., Zaritsky, D., McCarthy, D.~W., et al.\ 2005, \apj, 630, 206
\bibitem[Finn et al.(2008)]{2008ApJ...679..279F} Finn, R.~A., Balogh, M.~L., Zaritsky, D., et al.\ 2008, \apj, 679, 279
\bibitem[Fossati et al.(2016)]{2016MNRAS.455.2028F} Fossati, M., Fumagalli, M., Boselli, A., et al.\ 2016, \mnras, 455, 2028 
\bibitem[Fossati et al.(2017)]{2017ApJ...835..153F} Fossati, M., Wilman, D.~J., Mendel, J.~T., et al.\ 2017, \apj, 835, 153 
\bibitem[Fossati et al.(2018)]{2018A&A...614A..57F} Fossati, M., Mendel, J.~T., Boselli, A., et al.\ 2018, \aap, 614, A57 
\bibitem[Fossati et al.(2021)]{2021MNRAS.503.3044F} Fossati, M., Fumagalli, M., Lofthouse, E.~K., et al.\ 2021, \mnras, 503, 3044
\bibitem[Fujita et al.(2003)]{2003ApJ...586L.115F} Fujita, S.~S., Ajiki, M., Shioya, Y., et al.\ 2003, \apjl, 586, L115
\bibitem[Fumagalli et al.(2009)]{2009ApJ...697.1811F} Fumagalli, M., Krumholz, M.~R., Prochaska, J.~X., et al.\ 2009, \apj, 697, 1811
\bibitem[Fumagalli et al.(2011)]{2011ApJ...741L..26F} Fumagalli, M., da Silva, R.~L., \& Krumholz, M.~R.\ 2011a, \apjl, 741, L26 
\bibitem[Fumagalli et al.(2014)]{2014MNRAS.445.4335F} Fumagalli, M., Fossati, M., Hau, G.~K.~T., et al.\ 2014, \mnras, 445, 4335 
\bibitem[Gallego et al.(1995)]{1995ApJ...455L...1G} Gallego, J., Zamorano, J., Aragon-Salamanca, A., et al.\ 1995, \apjl, 455, L1
\bibitem[Gavazzi et al.(1998)]{1998AJ....115.1745G} Gavazzi, G., Catinella, B., Carrasco, L., Boselli, A., \& Contursi, A.\ 1998, \aj, 115, 1745 
\bibitem[Gavazzi et al.(1999)]{1999MNRAS.304..595G} Gavazzi, G., Boselli, A., Scodeggio, M., Pierini, D., \& Belsole, E.\ 1999, \mnras, 304, 595 
\bibitem[Gavazzi et al.(2003)]{2003A&A...400..451G} Gavazzi, G., Boselli, A., Donati, A., Franzetti, P., \& Scodeggio, M.\ 2003, \aap, 400, 451 
\bibitem[Gavazzi et al.(2004)]{2004A&A...417..499G} Gavazzi, G., Zaccardo, A., Sanvito, G., Boselli, A., \& Bonfanti, C.\ 2004, \aap, 417, 499 
\bibitem[Gavazzi et al.(2005)]{2005A&A...429..439G} Gavazzi, G., Boselli, A., van Driel, W., \& O'Neil, K.\ 2005, \aap, 429, 439 
\bibitem[Gavazzi et al.(2010)]{2010A&A...517A..73G} Gavazzi, G., Fumagalli, M., Cucciati, O., et al.\ 2010, \aap, 517, A73
\bibitem[Gilbank \& Balogh(2008)]{2008MNRAS.385L.116G} Gilbank, D.~G. \& Balogh, M.~L.\ 2008, \mnras, 385, L116
\bibitem[Giovanelli et al.(2005)]{2005AJ....130.2598G} Giovanelli, R., Haynes, M.~P., Kent, B.~R., et al.\ 2005, \aj, 130, 2598 
\bibitem[Gnedin(2003)]{2003ApJ...589..752G} Gnedin, O.~Y.\ 2003, \apj, 589, 752 
\bibitem[G{\'o}mez et al.(2003)]{2003ApJ...584..210G} G{\'o}mez, P.~L., Nichol, R.~C., Miller, C.~J., et al.\ 2003, \apj, 584, 210 
\bibitem[Gullieuszik et al.(2017)]{2017ApJ...846...27G} Gullieuszik, M., Poggianti, B.~M., Moretti, A., et al.\ 2017, \apj, 846, 27
\bibitem[Gunawardhana et al.(2013)]{2013MNRAS.433.2764G} Gunawardhana, M.~L.~P., Hopkins, A.~M., Bland-Hawthorn, J., et al.\ 2013, \mnras, 433, 2764 
\bibitem[Gunn \& Gott(1972)]{1972ApJ...176....1G} Gunn, J.~E., \& Gott, J.~R., III 1972, \apj, 176, 1 
\bibitem[Gwyn(2008)]{2008PASP..120..212G} Gwyn, S.~D.~J.\ 2008, \pasp, 120, 212 
\bibitem[Haynes \& Giovanelli(1984)]{1984AJ.....89..758H} Haynes, M.~P. \& Giovanelli, R.\ 1984, \aj, 89, 758
\bibitem[Hippelein et al.(2003)]{2003A&A...402...65H} Hippelein, H., Maier, C., Meisenheimer, K., et al.\ 2003, \aap, 402, 65 
\bibitem[Iglesias-P{\'a}ramo et al.(2002)]{2002A&A...384..383I} Iglesias-P{\'a}ramo, J., Boselli, A., Cortese, L., et al.\ 2002, \aap, 384, 383
\bibitem[Iglesias-P{\'a}ramo et al.(2003)]{2003A&A...397..421I} Iglesias-P{\'a}ramo, J., Boselli, A., Gavazzi, G., et al.\ 2003, \aap, 397, 421
\bibitem[Izotov et al.(2016)]{2016MNRAS.461.3683I} Izotov, Y.~I., Schaerer, D., Thuan, T.~X., et al.\ 2016, \mnras, 461, 3683
\bibitem[J{\'a}chym et al.(2014)]{2014ApJ...792...11J} J{\'a}chym, P., Combes, F., Cortese, L., Sun, M., \& Kenney, J.~D.~P.\ 2014, \apj, 792, 11 
\bibitem[Jones \& Bland-Hawthorn(2001)]{2001ApJ...550..593J} Jones, D.~H. \& Bland-Hawthorn, J.\ 2001, \apj, 550, 593
\bibitem[Joshi et al.(2020)]{2020MNRAS.496.2673J} Joshi, G.~D., Pillepich, A., Nelson, D., et al.\ 2020, \mnras, 496, 2673
\bibitem[Joshi et al.(2021)]{2021MNRAS.508.1652J} Joshi, G.~D., Pillepich, A., Nelson, D., et al.\ 2021, \mnras, 508, 1652
\bibitem[Junais et al.(2021)]{2021A&A...650A..99J} Junais, Boissier, S., Boselli, A., et al.\ 2021, \aap, 650, A99
\bibitem[Junais et al.(2022)]{2022A&A...667A..76J} Junais, Boissier, S., Boselli, A., et al.\ 2022, \aap, 667, A76
\bibitem[Kenney et al.(2014)]{2014ApJ...780..119K} Kenney, J.~D.~P., Geha, M., J{\'a}chym, P., et al.\ 2014, \apj, 780, 119 
\bibitem[Kennicutt(1998)]{1998ARA&A..36..189K} Kennicutt, R.~C., Jr.\ 1998a, \araa, 36, 189 
\bibitem[Kodama et al.(2004)]{2004MNRAS.354.1103K} Kodama, T., Balogh, M.~L., Smail, I., et al.\ 2004, \mnras, 354, 1103
\bibitem[Koyama et al.(2010)]{2010MNRAS.403.1611K} Koyama, Y., Kodama, T., Shimasaku, K., et al.\ 2010, \mnras, 403, 1611
\bibitem[Koyama et al.(2011)]{2011ApJ...734...66K} Koyama, Y., Kodama, T., Nakata, F., et al.\ 2011, \apj, 734, 66
\bibitem[Koyama et al.(2018)]{2018PASJ...70S..21K} Koyama, Y., Hayashi, M., Tanaka, M., et al.\ 2018, \pasj, 70, S21
\bibitem[Lake et al.(1998)]{1998ApJ...495..152L} Lake, G., Katz, N., \& Moore, B.\ 1998, \apj, 495, 152
\bibitem[Larson et al.(1980)]{1980ApJ...237..692L} Larson, R.~B., Tinsley, B.~M., \& Caldwell, C.~N.\ 1980, \apj, 237, 692 
\bibitem[Lewis et al.(2002)]{2002MNRAS.334..673L} Lewis, I., Balogh, M., De Propris, R., et al.\ 2002, \mnras, 334, 673 
\bibitem[Lieder et al.(2012)]{2012A&A...538A..69L} Lieder, S., Lisker, T., Hilker, M., et al.\ 2012, \aap, 538, A69
\bibitem[Lim et al.(2020)]{2020ApJ...899...69L} Lim, S., C{\^o}t{\'e}, P., Peng, E.~W., et al.\ 2020, \apj, 899, 69
\bibitem[Liu et al.(2021)]{2021ApJ...908..228L} Liu, Q., Yee, H.~K.~C., Drissen, L., et al.\ 2021, \apj, 908, 228.
\bibitem[Longobardi et al.(2020)]{2020A&A...644A.161L} Longobardi, A., Boselli, A., Fossati, M., et al.\ 2020, \aap, 644, A161
\bibitem[Lotz et al.(2019)]{2019MNRAS.488.5370L} Lotz, M., Remus, R.-S., Dolag, K., et al.\ 2019, \mnras, 488, 5370
\bibitem[Ly et al.(2007)]{2007ApJ...657..738L} Ly, C., Malkan, M.~A., Kashikawa, N., et al.\ 2007, \apj, 657, 738 
\bibitem[Marinacci et al.(2018)]{2018MNRAS.480.5113M} Marinacci, F., Vogelsberger, M., Pakmor, R., et al.\ 2018, \mnras, 480, 5113
\bibitem[Mateo(1998)]{1998ARA&A..36..435M} Mateo, M.~L.\ 1998, \araa, 36, 435
\bibitem[Matharu et al.(2021)]{2021ApJ...923..222M} Matharu, J., Muzzin, A., Brammer, G.~B., et al.\ 2021, \apj, 923, 222.
\bibitem[McGee et al.(2009)]{2009MNRAS.400..937M} McGee, S.~L., Balogh, M.~L., Bower, R.~G., Font, A.~S., \& McCarthy, I.~G.\ 2009, \mnras, 400, 937 
\bibitem[Mehta et al.(2015)]{2015ApJ...811..141M} Mehta, V., Scarlata, C., Colbert, J.~W., et al.\ 2015, \apj, 811, 141
\bibitem[Mei et al.(2007)]{2007ApJ...655..144M} Mei, S., Blakeslee, J.~P., C{\^o}t{\'e}, P., et al.\ 2007, \apj, 655, 144 
\bibitem[Melekh et al.(2015)]{2015MNRAS.450..111M} Melekh, B., Recchi, S., Hensler, G., et al.\ 2015, \mnras, 450, 111
\bibitem[Merluzzi et al.(2013)]{2013MNRAS.429.1747M} Merluzzi, P., Busarello, G., Dopita, M.~A., et al.\ 2013, \mnras, 429, 1747
\bibitem[Merritt(1983)]{1983ApJ...264...24M} Merritt, D.\ 1983, \apj, 264, 24 
\bibitem[Moore et al.(1998)]{1998ApJ...495..139M} Moore, B., Lake, G., \& Katz, N.\ 1998, \apj, 495, 139 
\bibitem[Morioka et al.(2008)]{2008PASJ...60.1219M} Morioka, T., Nakajima, A., Taniguchi, Y., et al.\ 2008, \pasj, 60, 1219
\bibitem[Moss \& Whittle(2000)]{2000MNRAS.317..667M} Moss, C., \& Whittle, M.\ 2000, \mnras, 317, 667 
\bibitem[Moss et al.(1988)]{1988MNRAS.232..381M} Moss, C., Whittle, M., \& Irwin, M.~J.\ 1988, \mnras, 232, 381
\bibitem[Naiman et al.(2018)]{2018MNRAS.477.1206N} Naiman, J.~P., Pillepich, A., Springel, V., et al.\ 2018, \mnras, 477, 1206
\bibitem[Nantais et al.(2020)]{2020MNRAS.499.3061N} Nantais, J., Wilson, G., Muzzin, A., et al.\ 2020, \mnras, 499, 3061
\bibitem[Nelson et al.(2018)]{2018MNRAS.475..624N} Nelson, D., Pillepich, A., Springel, V., et al.\ 2018, \mnras, 475, 624
\bibitem[Nelson et al.(2019)]{2019ComAC...6....2N} Nelson, D., Springel, V., Pillepich, A., et al.\ 2019a, Computational Astrophysics and Cosmology, 6, 2
\bibitem[Nelson et al.(2019)]{2019MNRAS.490.3234N} Nelson, D., Pillepich, A., Springel, V., et al.\ 2019b, \mnras, 490, 3234
\bibitem[Nulsen(1982)]{1982MNRAS.198.1007N} Nulsen, P.~E.~J.\ 1982, \mnras, 198, 1007 
\bibitem[Peng et al.(2010)]{2010ApJ...721..193P} Peng, Y.-j., Lilly, S.~J., Kova{\v c}, K., et al.\ 2010, \apj, 721, 193 
\bibitem[P{\'e}rez-Gonz{\'a}lez et al.(2003)]{2003ApJ...591..827P} P{\'e}rez-Gonz{\'a}lez, P.~G., Zamorano, J., Gallego, J., et al.\ 2003, \apj, 591, 827
\bibitem[Pillepich et al.(2018)]{2018MNRAS.473.4077P} Pillepich, A., Springel, V., Nelson, D., et al.\ 2018a, \mnras, 473, 4077
\bibitem[Pillepich et al.(2018)]{2018MNRAS.475..648P} Pillepich, A., Nelson, D., Hernquist, L., et al.\ 2018b, \mnras, 475, 648
\bibitem[Pillepich et al.(2019)]{2019MNRAS.490.3196P} Pillepich, A., Nelson, D., Springel, V., et al.\ 2019, \mnras, 490, 3196
\bibitem[Poggianti et al.(2017)]{2017ApJ...844...48P} Poggianti, B.~M., Moretti, A., Gullieuszik, M., et al.\ 2017, \apj, 844, 48 
\bibitem[Popesso et al.(2005)]{2005A&A...433..415P} Popesso, P., B{\"o}hringer, H., Romaniello, M., et al.\ 2005, \aap, 433, 415
\bibitem[Quilis et al.(2000)]{2000Sci...288.1617Q} Quilis, V., Moore, B., \& Bower, R.\ 2000, Science, 288, 1617
\bibitem[Ramatsoku et al.(2019)]{2019MNRAS.487.4580R} Ramatsoku, M., Serra, P., Poggianti, B.~M., et al.\ 2019, \mnras, 487, 4580
\bibitem[Rines \& Geller(2008)]{2008AJ....135.1837R} Rines, K. \& Geller, M.~J.\ 2008, \aj, 135, 1837
\bibitem[Roediger \& Br{\"u}ggen(2007)]{2007MNRAS.380.1399R} Roediger, E., \& Br{\"u}ggen, M.\ 2007, \mnras, 380, 1399 
\bibitem[Sabatini et al.(2003)]{2003MNRAS.341..981S} Sabatini, S., Davies, J., Scaramella, R., et al.\ 2003, \mnras, 341, 981
\bibitem[Sandage et al.(1985)]{1985AJ.....90.1759S} Sandage, A., Binggeli, B., \& Tammann, G.~A.\ 1985, \aj, 90, 1759
\bibitem[Sander \& Hensler(2021)]{2021MNRAS.501.5330S} Sander, B. \& Hensler, G.\ 2021, \mnras, 501, 5330
\bibitem[Sardaneta et al.(2022)]{2022A&A...659A..45S} Sardaneta, M.~M., Amram, P., Boselli, A., et al.\ 2022, \aap, 659, A45
\bibitem[Schechter(1976)]{1976ApJ...203..297S} Schechter, P.\ 1976, \apj, 203, 297
\bibitem[Schlegel et al.(1998)]{1998ApJ...500..525S} Schlegel, D.~J., Finkbeiner, D.~P., \& Davis, M.\ 1998, \apj, 500, 525 
\bibitem[Schulz \& Struck(2001)]{2001MNRAS.328..185S} Schulz, S. \& Struck, C.\ 2001, \mnras, 328, 185
\bibitem[Shimakawa et al.(2018)]{2018MNRAS.473.1977S} Shimakawa, R., Kodama, T., Hayashi, M., et al.\ 2018, \mnras, 473, 1977
\bibitem[Shioya et al.(2008)]{2008ApJS..175..128S} Shioya, Y., Taniguchi, Y., Sasaki, S.~S., et al.\ 2008, \apjs, 175, 128
\bibitem[Sobral et al.(2011)]{2011MNRAS.411..675S} Sobral, D., Best, P.~N., Smail, I., et al.\ 2011, \mnras, 411, 675
\bibitem[Sobral et al.(2016)]{2016MNRAS.458.3443S} Sobral, D., Stroe, A., Koyama, Y., et al.\ 2016, \mnras, 458, 3443
\bibitem[Solanes et al.(2001)]{2001ApJ...548...97S} Solanes, J.~M., Manrique, A., Garc{\'{\i}}a-G{\'o}mez, C., et al.\ 2001, \apj, 548, 97 
\bibitem[Solanes et al.(2002)]{2002AJ....124.2440S} Solanes, J.~M., Sanchis, T., Salvador-Sol{\'e}, E., et al.\ 2002, \aj, 124, 2440
\bibitem[Somerville \& Dav{\'e}(2015)]{2015ARA&A..53...51S} Somerville, R.~S. \& Dav{\'e}, R.\ 2015, \araa, 53, 51
\bibitem[Springel et al.(2018)]{2018MNRAS.475..676S} Springel, V., Pakmor, R., Pillepich, A., et al.\ 2018, \mnras, 475, 676
\bibitem[Sternberg et al.(2003)]{2003ApJ...599.1333S} Sternberg, A., Hoffmann, T.~L., \& Pauldrach, A.~W.~A.\ 2003, \apj, 599, 1333
\bibitem[Stevens et al.(2019)]{2019MNRAS.483.5334S} Stevens, A.~R.~H., Diemer, B., Lagos, C. del P., et al.\ 2019, \mnras, 483, 5334
\bibitem[Stott et al.(2007)]{2007ApJ...661...95S} Stott, J.~P., Smail, I., Edge, A.~C., et al.\ 2007, \apj, 661, 95
\bibitem[Stott et al.(2009)]{2009MNRAS.394.2098S} Stott, J.~P., Pimbblet, K.~A., Edge, A.~C., et al.\ 2009, \mnras, 394, 2098
\bibitem[Stroe et al.(2017)]{2017MNRAS.465.2916S} Stroe, A., Sobral, D., Paulino-Afonso, A., et al.\ 2017, \mnras, 465, 2916
\bibitem[Sun et al.(2006)]{2006ApJ...637L..81S} Sun, M., Jones, C., Forman, W., et al.\ 2006, \apjl, 637, L81 
\bibitem[Tonnesen et al.(2007)]{2007ApJ...671.1434T} Tonnesen, S., Bryan, G.~L., \& van Gorkom, J.~H.\ 2007, \apj, 671, 1434
\bibitem[Trentham \& Hodgkin(2002)]{2002MNRAS.333..423T} Trentham, N. \& Hodgkin, S.\ 2002, \mnras, 333, 423
\bibitem[Tresse \& Maddox(1998)]{1998ApJ...495..691T} Tresse, L. \& Maddox, S.~J.\ 1998, \apj, 495, 691
\bibitem[Umeda et al.(2004)]{2004ApJ...601..805U} Umeda, K., Yagi, M., Yamada, S.~F., et al.\ 2004, \apj, 601, 805
\bibitem[van der Wel et al.(2014)]{2014ApJ...788...28V} van der Wel, A., Franx, M., van Dokkum, P.~G., et al.\ 2014, \apj, 788, 28
\bibitem[Vollmer et al.(2001)]{2001ApJ...561..708V} Vollmer, B., Cayatte, V., Balkowski, C., \& Duschl, W.~J.\ 2001a, \apj, 561, 708 
\bibitem[Vollmer et al.(2004)]{2004A&A...419...35V} Vollmer, B., Balkowski, C., Cayatte, V., van Driel, W., \& Huchtmeier, W.\ 2004, \aap, 419, 35 
\bibitem[Vollmer et al.(2008)]{2008A&A...491..455V} Vollmer, B., Braine, J., Pappalardo, C., \& Hily-Blant, P.\ 2008, \aap, 491, 455 
\bibitem[Vollmer et al.(2021)]{2021A&A...645A.121V} Vollmer, B., Fossati, M., Boselli, A., et al.\ 2021, \aap, 645, A121
\bibitem[Vulcani et al.(2017)]{2017ApJ...837..126V} Vulcani, B., Treu, T., Nipoti, C., et al.\ 2017, \apj, 837, 126
\bibitem[Yagi et al.(2002)]{2002AJ....123...87Y} Yagi, M., Kashikawa, N., Sekiguchi, M., et al.\ 2002, \aj, 123, 87
\bibitem[Yagi et al.(2010)]{2010AJ....140.1814Y} Yagi, M., Yoshida, M., Komiyama, Y., et al.\ 2010, \aj, 140, 1814 
\bibitem[Yagi et al.(2017)]{2017ApJ...839...65Y} Yagi, M., Yoshida, M., Gavazzi, G., et al.\ 2017, \apj, 839, 65 
\bibitem[Westra et al.(2010)]{2010ApJ...708..534W} Westra, E., Geller, M.~J., Kurtz, M.~J., et al.\ 2010, \apj, 708, 534
\bibitem[Wetzel et al.(2013)]{2013MNRAS.432..336W} Wetzel, A.~R., Tinker, J.~L., Conroy, C., \& van den Bosch, F.~C.\ 2013, \mnras, 432, 336 
\bibitem[Wright et al.(2010)]{2010AJ....140.1868W} Wright, E.~L., Eisenhardt, P.~R.~M., Mainzer, A.~K., et al.\ 2010, \aj, 140, 1868-1881 
\bibitem[Zabel et al.(2022)]{2022ApJ...933...10Z} Zabel, N., Brown, T., Wilson, C.~D., et al.\ 2022, \apj, 933, 10
\bibitem[Zeimann et al.(2013)]{2013ApJ...779..137Z} Zeimann, G.~R., Stanford, S.~A., Brodwin, M., et al.\ 2013, \apj, 779, 137
\bibitem[Zhang et al.(2013)]{2013ApJ...777..122Z} Zhang, B., Sun, M., Ji, L., et al.\ 2013, \apj, 777, 122 

\end{thebibliography}
\end{document}